\documentclass{aa}
\usepackage{natbib}
\usepackage{graphicx}
\usepackage{txfonts}
\def\EBV{\mbox{E$_{\rm B-V}$}}

\def\AV{\mbox{A$_{\rm V}$}}

\def\HH{\mbox{H$_2$}}

\def\nH2{{\rm n}({\rm H}_2)}
\def\NH2{{\rm N}({\rm H}_2)}
 
\def\pccc{~{\rm cm}^{-3}} 
\def\pcc {~{\rm cm}^{-2}}

\def\Tstar {\mbox{${\rm T}_{\rm R}^*$}}
\def\Tsub#1 {\mbox{$T_{\rm #1}$}}
\def\TK  {\Tsub K }
\def\TB  {\Tsub B }

\def\Texc {\Tsub exc }

\def\Tcmb{\Tsub cmb }

\def\arcsec{\mbox{$^{\prime\prime}$}} \def\arcmin{\mbox{$^{\prime}$}}

\def\degr{\mbox{$^{\rm o}$}}

\def\p{\mbox{$^+$}}
\def\cotw {\mbox{$^{12}$CO}}
\def\coth {\mbox{$^{13}$CO}}

\def\hcop{\mbox{{HCO\p}}}

\def\cch{\mbox{C$_2$H}}
  
\def\hhco{\mbox{H$_2$CO}}
\def\h13cop{\mbox{{H$^{13}$CO\p}}}

\def\c3h2{\mbox{C$_3$H$_2$}}
 
 \def\R0{R$_0$} 
  \def\kpc{\rm kpc}
  
\def\ddeg{{}^\circ\kern-.1em}

\def\kms{\mbox{km\,s$^{-1}$}}

\def\bll{BL Lac}

\def\E#1 {$10^{#1}$}
\def\E#1 {E{#1}}
\def\P#1,{$\nH2\TK~=~#1\times~10^4\pccc$~K}
\def\ec#1,#2,#3,{#1\,(#2)\E{#3}}

\def\zoph{$\zeta$ Oph}

\def\H3{\mbox{H$_3$}}

\def\RH2{\mbox {R$_{\rm G}$}}
\def\fH2{\mbox {f$_{{\rm H}_2}$}}
\def\W13{\mbox{W$_{13}$}}
\def\WCO{\mbox{W$_{\rm CO}$}}
\def\FH2{\mbox {F$_{\HH}$}}
\def\W13{\mbox{W$_{^{13}{\rm CO}}$}}
\def\WCO{\mbox{W$_{\rm CO}$}}

\newcommand{\emm}[1]{\ensuremath{#1}}   
\newcommand{\emr}[1]{\emm{\mathrm{#1}}} 

\newcommand{\mean}[1]{\emm{ \left<  #1 \right> }}

\newcommand{\twCO}{\emr{^{12}CO}}

\newcommand{\Kkms}{\emr{\,K\,km\,s^{-1}}}

\title{Imaging diffuse clouds: Bright and dark gas mapped in CO
  \thanks{Based on observations obtained with the ARO Kitt Peak 12m
    telescope.}}

\author{H. S. Liszt\inst{1}, J. Pety\inst{2,3}}
\institute{National Radio Astronomy Observatory, 520 Edgemont Road,
  Charlottesville, VA, USA 22903-2475 \and Institut de Radioastronomie
  Millim\'etrique, 300 Rue de la Piscine, F-38406 Saint Martin d'H\`eres,
  France \and Obs. de Paris, 61 av. de l'Observatoire, 75014, Paris, France
}

\begin{document}
\date{received \today}%
\offprints{H. S. Liszt}%
\mail{hliszt@nrao.edu}%

\abstract
{}
{We wish to relate the degree scale structure of galactic diffuse clouds to
  sub-arcsecond atomic and molecular absorption spectra obtained against
  extragalactic continuum background sources.}
{We used the ARO 12m telescope to map J=1-0 CO emission at 1\arcmin\ 
  resolution over 30\arcmin\ fields around the positions of 11 background
  sources occulted by 20 molecular absorption line components, of which 11
  had CO emission counterparts.  We compare maps of CO emission to
  sub-arcsec atomic and molecular absorption spectra and to the large-scale
  distribution of interstellar reddening.}
{ 1) The same clouds, identified by their velocity, were seen in absorption
  and emission and atomic and molecular phases, not necessarily in the same
  direction.  Sub-arcsecond absorption spectra are a preview of what is
  seen in CO emission away from the continuum.  2) The CO emission
  structure was amorphous in 9 cases, quasi-periodic or wave-like around
  B0528+134 and tangled and filamentary around \bll.  3) Strong emission,
  typically 4-5 K at \EBV\ $\le$ 0.15 mag and up to 10-12 K at \EBV\ $\la$
  0.3 mag was found, much brighter than toward the background targets.
  Typical covering factors of individual features at the 1 \Kkms\ level
  were 20\%.  4) CO-\HH\ conversion factors as much as 4-5 times below the
  mean value N(\HH)/\WCO\ = $2\times 10^{20}~\HH \pcc$(\Kkms)$^{-1}$ are
  required to explain the luminosity of CO emission at/above the level of
  1\Kkms.  Small conversion factors and sharp variability of the conversion
  factor on arcminute scales are due primarily to CO chemistry and need not
  represent unresolved variations in reddening or total column density.}
{ Like FERMI and PLANCK we see some gas that is dark in CO and other gas in
  which CO is overluminous per \HH.  A standard CO-\HH{} conversion factor
  applies overall owing to balance between the luminosities per \HH\ and
  surface covering factors of bright and dark CO, but with wide variations
  between sightlines and across the faces of individual clouds.  }

\keywords{ interstellar medium -- molecules }
  
\maketitle{}

\section{Introduction}

With somewhat imprecise boundaries, interstellar clouds are generally
classed as diffuse, \AV $\la$ 1 mag, or dark, \AV $\ga$ 4-6 mag, with an
intermediate translucent regime \citep{SnoMcC06}.  In diffuse clouds the
dominant form of carbon is C\p\ and hydrogen is mostly atomic, although
with a very significant overall admixture of \HH, 25\% or more as a global
average \citep{SavDra+77,LisPet+10}.  In dark or molecular clouds the
carbon is overwhelmingly in CO with an admixture of C I and the hydrogen
resides almost entirely in \HH.  The population of diffuse clouds is
sometimes called H I clouds in radio astronomical terms.

The shadows of dark clouds are seen outlined against brighter background
fields and the clouds themselves are often imaged in the mm-wave emission
of CO and many other species: the {\twCO{}(1-0) sky \citep{DamHar+01} is
  usually (and in part incorrectly, see below) understood as a map of
  fully-molecular clouds.  The shadows of H I or diffuse clouds are their
  absorption-line spectra and for the most part, individual diffuse clouds
  are known only as kinematic features in optical and/or radio absorption
  spectra.  No means exist to image individual diffuse clouds at optical
  wavelengths and attempts to map individual H I clouds at radio
  wavelengths are generally frustrated by the blending and overlapping of
  contributions from multiple clouds and gas phases.  This lack of identity
  has greatly complicated our ability to define diffuse clouds physically
  because absorption lines do not generally permit a direct determination
  of the cloud size or internal density.
  
  When diffuse clouds discovered in absorption-line spectra have a
  sufficiently high complement of molecules they may be imaged at radio
  wavelengths in species such as OH and CH and, most usefully, CO.  Despite
  a low fractional abundance of CO relative to \HH, $\mean{{\rm X(CO)}} =
  3\times 10^{-6}$ \citep{BurFra+07}, mapping is facilitated by an enhanced
  brightness of the J=1-0 line in diffuse gas: the temperature is somewhat
  elevated (\TK $\ga 25$ K), the density is comparatively small at typical
  ambient thermal pressure \citep{JenTri11} and the rotation ladder is
  subthermally excited.  In accord with theory \citep{GolKwa74}, it is
  found observationally that there is a simple, linear proportionality
  between the integrated intensity \WCO\ of the CO J=1-0 lines and the CO
  column density, even when the gas is optically thick: N(CO) $\approx
  10^{15}\pcc$ \WCO/\Kkms for \WCO\ $\approx 0.2 - 6$ \Kkms
  \citep{LisLuc98,Lis07CO}.  Per molecule, the ratio \WCO/N(CO) is 30-50
  times higher in diffuse gas, compared to conditions in dense shielded
  fully-molecular gas where the rotation ladder is thermalized
  \citep{LisPet+10}.  Of course this is of substantial assistance in the
  present work.  Conversely, the high brightness (5-12 K) of many of lines
  we detected should \emph{not} be taken as discrediting their origin in
  diffuse gas.

\begin{figure*}
  \includegraphics[height=11.75cm]{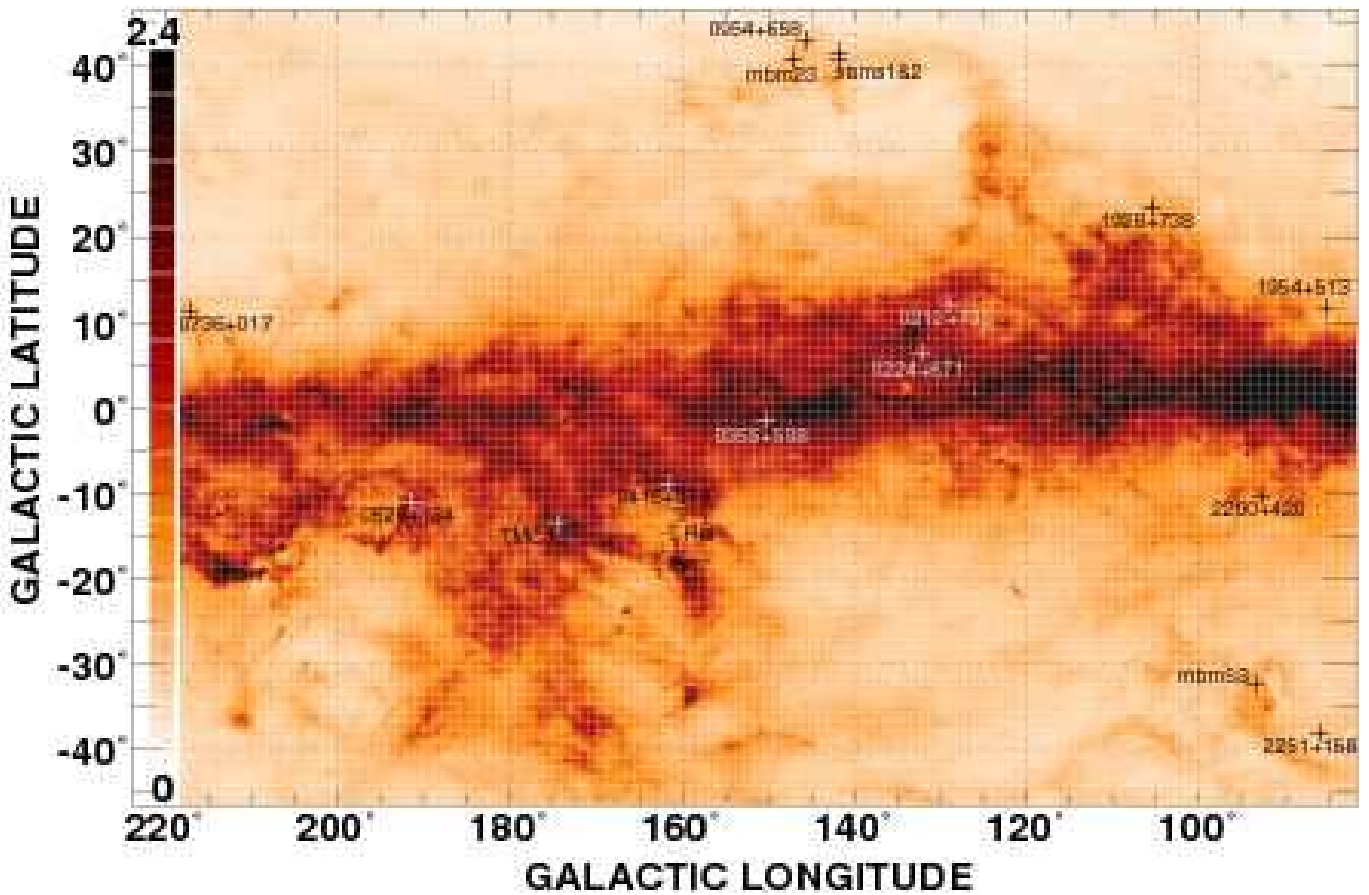}
\caption[]{Outer-galaxy finding chart for the sources studied here except 
  B1730-130 at $l = 12$\degr, see Table 1.  The colored background is
  reddening at 6\arcmin\ resolution \citep{SchFin+98} truncated at a
  maximum of 2.4 mag as shown on the bar scale at left.  Positions of
  continuum sources are indicated (see Table 1) along with a few other
  objects: the high-latitude molecular clouds MBM23 and MBM 53
  \citep{MagBli+85}; TMC-1; SAMS 1 and 2 \citep{Hei04}; and two Perseus
  stars commonly used for optical absorption line studies.}
\end{figure*}

\begin{table*}
\caption[]{Continuum target, line of sight and map field properties$^1$}
{
\small
\begin{tabular}{lccccccccccc}
\hline
Target & ra & dec & l  &  b &  Map & \EBV$^2$ & N (H I)$^3$ & N(\HH)$^4$ & \fH2$^7$ & \WCO & $<$\WCO$>$ \\
       & (J2000)  & (J2000) & & & size & mag   & $10^{20}\pcc$ & $10^{20}\pcc$ & & \Kkms & \Kkms\\
\hline
B0736+017 & 07:39:18.03 & 01:37:04.6 &  216.99 & 11.38 & 15\arcmin &0.13 & 7.7  & 3.3 & 0.46 & 0.8 & 0.4\\
B0954+658 & 09:58:47.24 & 65:33:54.7 & 145.75 & 43.13 & 30\arcmin & 0.12 & 5.3$^5$  & 4.8 & 0.64 & 1.6 & 0.6\\
B1730-130 & 17:33:02.66 & -13:04:49.5 & 12.03  & 10.81 & 30\arcmin & 0.53 & 28.3 & 4.3 & 0.23 & 0.4 & 0.4\\
B1928+738 & 19:27:48.58 & 73:58:01.6 & 105.63 & 23.54 & 20\arcmin & 0.13 & 7.2$^5$ & 2.7 & 0.43 & $<$0.2 & 0.0 \\
B1954+513 & 19:55:42.69 & 51.31:48.5 & 85.30 & 11.76 & 30\arcmin & 0.15 & 12.7$^5$ & 5.0 & 0.44 & 1.6 & 0.3 \\
B2251+158 & 22:53:57.71 & 16:08:53.4 & 86.11 & -38.18 & 30\arcmin &0.10 & 4.6 & 1.2 & 0.34 & 0.8 & 0.2 \\ 
\hline
B0528+134 & 05:30:46.41 & 13:31:55.1 & 191.37 & -11.01 & 30\arcmin & 0.89 & 30.9 & 7.9 & 0.34 & 2.2 & 2.6\\
B2200+420 & 22:02:43.24 & 42:16:39.9 & 92.59 & -10.44 & 30\arcmin & 0.33 & 9.7 & 8.8 & 0.66 & 5.8 & 3.7\\
\hline
B0212+735 & 02:17:30.81 & 73:49:32.6 & 128.93& 11.96 & 30\arcmin & 0.76 & 32.1 & 18.6 & 0.54 & 5.8 & 1.56\\
B0224+671 & 02:28:50.03 & 67:21:31.3 &132.12 & 6.23 & 30\arcmin &1.00 & 38.3 & 9.2 & 0.32 &1.9 & 4.0 \\
B0355+508 & 03:59:29.73 & 50:57:50.1 &150.38 &-1.60 &30\arcmin x 50\arcmin & (1.50)$^6$ & 111.3 & 24.2 & 0.30 
& 14.3 & 4.2\\
\hline
Mean & & & & & & 0.41 & 17.7 & 6.58 & 0.43 &  2.09 & 1.38 \\
\hline
\end{tabular}}
\\
$^1$Sources are placed in three groups according to their discussion in Sect. 3, 4 and 5 \\
$^2$from \cite{SchFin+98} \\
$^3$ N(H I) = $2.6 \times 10^{20}\pcc \int{\tau(H I)}dv$ (see Sect 2) except where noted \\
$^4$ N(\HH) = N(\hcop)/$3\times10^{-9}$ see Sect. 2. \\
$^5$ from \cite{HarBur97}, N(H I) = $1.823 \times 10^{18}\pcc \int{\TB(H I)}dv$  \\
$^6$ at such a low galactic latitude \EBV\ is not reliably determined \\
$^7$ \fH2 = 2N(\HH)/(N(H I) + 2N(\HH))
\\
\end{table*}

Earlier we showed that, in the mean, CO-\HH\ conversion factors are similar
in diffuse and dense fully molecular gas \citep{LisPet+10}, because the
small abundance of CO relative to \HH\ in diffuse gas is compensated by a
much higher brightness per CO molecule.  But the proportionality between
\WCO\ and N(CO) in diffuse gas, where CO represents such a small fraction
of the available gas phase carbon, means that the CO map of a diffuse cloud
is really an image of the CO chemistry.  Moreover the CO abundance exhibits
extreme sensitivities to local conditions that are manifested as order of
magnitude scatter in N(CO)/N(\HH) in optical absorption line studies
\citep{SonWel+07,BurFra+07,SheRog+07,SheRog+08}, even beyond the
often-rapid variation of N(\HH) with \EBV\ \citep{SavDra+77} (\EBV\ 
$\approx$ \AV/3.1).  The net result is that the CO emission map of a
diffuse cloud can only indirectly be interpreted as tracing the underlying
mass distribution, or even that of the \HH.  Nonetheless, it should (we
hope) provide some impression of the nature of the host gas, especially in
the absence of any other means of ascertaining this.

In this paper we present maps of CO J=1-0 emission at arcminute resolution
over 11 sky fields, typically 30\arcmin $\times$ 30\arcmin\ around the
positions of compact extragalactic mm-wave continuum sources that we have
long used as targets for absorption line studies of the chemistry of
diffuse clouds.  As is the case for nearly all background sources seen at
galactic latitudes $|b| < 15-18$\degr, and for some sources at higher
latitudes, the current targets were known to show absorption from \hcop\ 
and from one or more other commonly-detected species (OH, CO, \cch, \c3h2);
most but not all directions also were known to show CO emission in at least
some of the kinematic features present in absorption.

This work is organized as follows. The observational material discussed
here is summarized in Sect. 2.  In Sects. 3-5 we discuss the new maps with
sources grouped in order of kinematic complexity.  Sect. 6 is an
intermediate summary of the lessons drawn from close scrutiny of the maps.
Sect.  7 briefly discusses the influences of galactic and internal cloud
kinematics and Sect. 8 presents a comparison of CO intensity and reddening
within a few of the simpler individual fields.  Online Appendix A shows a
few position-velocity diagrams that, while of interest, could be considered
redundant with those shown in the main text in Figs. 13 and 14.  Figures
B.1 and B.2 in online Appendix B show the target lines of sight in the
context of large-scale galactic kinematics sampled in H I emission.

\section{Observational material}

\subsection{CO J=1-0 emission}

On-the-fly maps of CO J=1-0 emission were made at the ARO 12m telescope in
2008 December, 2009 January and 2009 December in generally poor weather
using filter banks with 100 kHz or 0.260 \kms{} channel spacing and
spectral resolution.  System temperatures were typically $450 - 750$ K.
The data were subsequently put onto 20\arcsec\ pixel grids using the AIPS
tasks OTFUV and SDGRD; the final spatial resolution is 1\arcmin.  Most maps
are approximately 30\arcmin\ $\times$ 30\arcmin\ on the sky and were
completed in 4-5 hours total observing time.  The new CO emission data are
presented in terms of the \Tstar\ scale in use at the 12m antenna and all
velocites are referred to the kinematic Local Standard of Rest.  The
typical rms channel-channel noise in these maps at 1\arcmin\ and 0.26 \kms\ 
resolution is 0.4-0.5 K; their sensitivity is rather moderate and the
detectability limit is of order 1 \Kkms{} for a single line component.

More sensitive CO J=1-0 line profiles at higher spectral resolution (25
kHz) had been previously observed toward the continuum sources as part of
our survey efforts, for instance see \cite{LisLuc98}.  It is these profiles
that are displayed in the Figures shown here representing emission in the
specific direction of the background target and used to calculate line
profile integrals as quoted in Table 1.

Many interstellar clouds lie at distances of about 150 pc from the Sun,
just outside the Local Bubble.  At this distance the 1\arcmin\ resolution
of our CO mapping corresponds to 0.041 pc.

\subsection{H I absorption and emission}

The $\lambda$ 21cm H I absorption spectra shown here are largely from the
work of \cite{DicKul+83} augmented by a few spectra taken at the VLA in
2005 May.  The spectral resolution of this data is 0.4 - 1.0 \kms.

Figures B.1 and B.2 of the online Appendix B show latitude-velocity
diagrams of H I emission drawn from the Leiden-Dwingeloo Survey of
\cite{HarBur97}.

\subsection{Molecular absorption}

Also shown here are spectra of $\lambda 18$cm OH absorption from
\cite{LisLuc96} and mm-wave absorption spectra of CO \citep{LisLuc98},
\hcop\ \citep{LucLis96} \cch\ \citep{LucLis00C2H} and \hhco\ 
\citep{LisLuc+06}.

\subsection{Reddening}

Maps of reddening were constructed from the results of \cite{SchFin+98}.
This dataset has 6\arcmin\ spatial resolution on a 2.5\arcmin\ pixel grid.
The stated single-pixel error is a percentage, 16\%, of the pixel value.
On average, 1 mag of reddening corresponds to a neutral gas column N(H) $=
5.8 \times 10^{21}\pcc$ \citep{SavDra+77}.

\subsection{Target fields}

The positions and other observational properties are summarized in Table 1
where the sources are grouped according to their order of presentation in
Sect. 3, 4 and 5.  The groups appear in order of increasing reddening and
gas column density and decreasing distance from the galactic plane.  The
line profile integrals \WCO\ quoted in Table 1 result from the more
sensitive earlier observations noted in Sect. 2.1.  The mean values quoted
for \WCO\ along individual sightlines are averages over the new map data
taken for this work.

Table 2 gives some pixel statistics about noise levels and spatial covering
factors as discussed in Sect. 8.

\subsection{Presentation of observations}

A finding chart including all sources except B1730-130 is shown in Fig. 1
where the locations of the background targets are shown on a large-scale
map of reddening, along with locations of a few other landmark objects as
noted in the figure caption.

Maps and spectra of the target fields and background sources are shown in
Figs.  2-12.  Within each of three groups, sources appear in order of
increasing right ascension. Members of the first group, shown in Figs. 2-7
and discussed in Sect. 3, are the simplest kinematically.  Figs. 8 and 9
show the fields around the background sources B0528+134 and B2200+420 (aka
\bll) that are also kinematically simple but are heavily patterned and
rather bright in CO emission; these are discussed in Sect. 4.  Figs. 10-12
(Sect. 5) show the results over three target fields with rather amorphous
structure whose kinematics are too complex to fit into the framework in
which the data for the other sources are presented in earlier figures.  Two
of these sources (B0212+735 and B0224+671) are relatively near each other
on the sky and sample similar galactic structure while the third target
B0355+508 (aka NRAO150) is the only source within 2\degr\ of the galactic
equator (see Table 1).

The format of Figures 2-11 is: at upper left a 90\arcmin\ map of \EBV\ from
the dataset of \cite{SchFin+98}, with an inset showing the field of view
mapped in CO, typically 30\arcmin\ on a side; at lower left a map of \WCO;
at lower right various atomic (H I) and molecular absorption spectra
showing the kinematic structure toward the background source; at upper
right, CO emission spectra of various sorts as depicted in the figure
captions.  The absorption spectra shown at lower right in these figures are
somewhat inhomogeneous because not all sources have the same full
complement of profiles.  In general, H I is at the bottom wherever possible
and above that are spectra of the most common molecules observed in
absorption; \hcop, observed toward all targets, OH, \cch\ and/or CO.  The
uppermost spectrum wherever possible is a species like \hhco\ or HNC
\citep{LisLuc+06,LisLuc01} that is detected less commonly and is indicative
of greater chemical complexity.

\begin{figure*}
  \includegraphics[height=13cm]{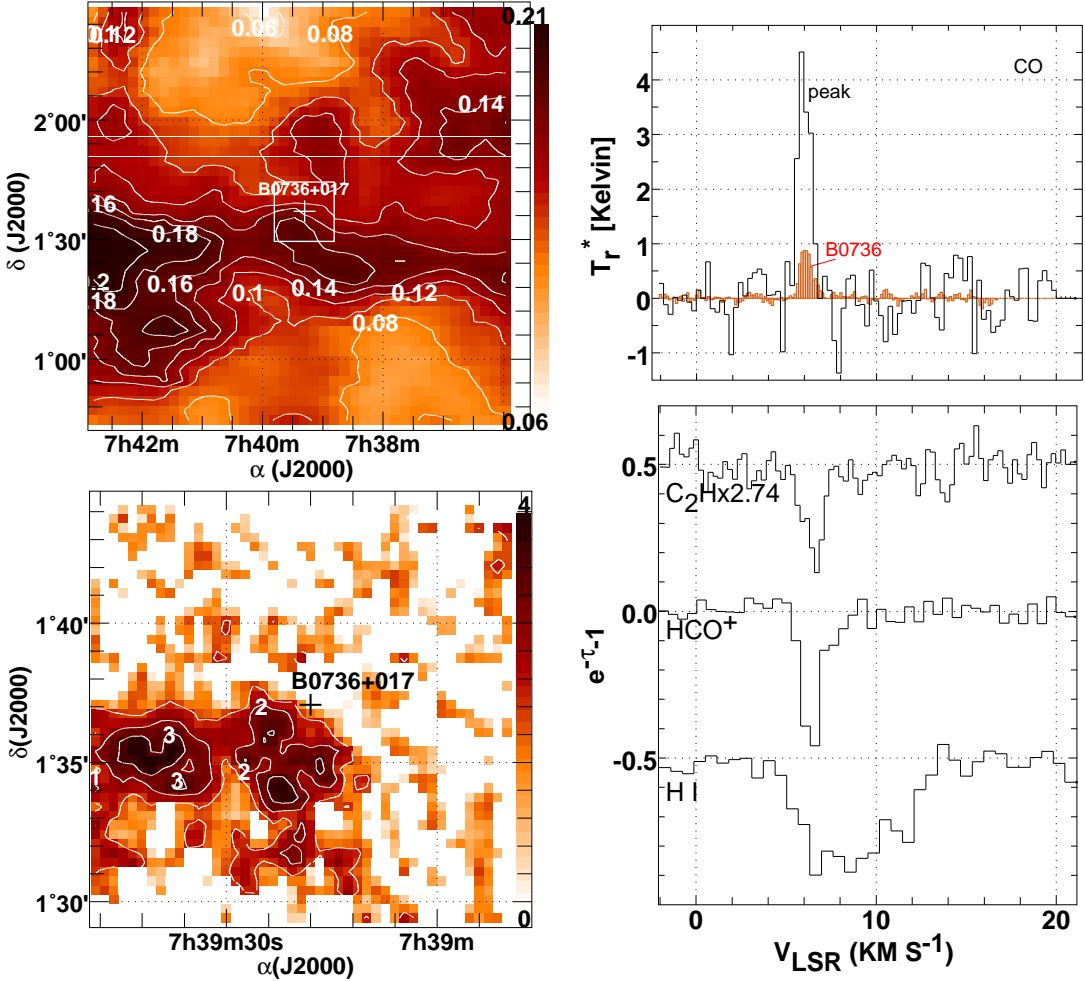}
\caption[]{The sky field around the position of B0736+017. Upper left:
  reddening at 6\arcmin\ resolution \citep{SchFin+98}.  Lower left:
  integrated CO emission \WCO\ in units of K \kms; contour levels are in
  1,2,... K \kms.  Lower right: absorption line profiles, scaled as noted.
  Upper right: {\bf CO emission,} toward B0736+017 and at the peak of the
  nearby CO distribution.}
\end{figure*}

Also shown for all sources are CO emission spectra at various locations in
the field mapped, as indicated in the spectra.  More complex aspects of the
presentation are discussed in the individual figure captions.

\begin{table}
\caption[]{Noise levels and spatial covering factors}
{
\small
\begin{tabular}{lccccc}
\hline
Target & V     & $\sigma_{\rm  profile}$ & $\sigma_{\rm map}$ & f$_{>1}^a$ & f$_{>2}^b$\\ 
       & \kms  & K  & \Kkms &  &  \\
\hline
B0736& 5.1,7.2 & 0.43 & 0.48 & 0.18 & 0.07 \\
B0954& 2.6,5.2 & 0.25 & 0.33 & 0.20 & 0.12 \\
B1730& 4.0,6.1 & 0.36 & 0.52 & 0.20 & 0.03 \\
B1928& -4.2,-0.8 & 0.46 & 0.52 & 0.03 & 0.00 \\
B1954& -1.0,2.4 & 0.33 & 0.35 & 0.20 & 0.13 \\
B2251&-10.8,-7.9 & 0.28 & 0.32 & 0.06 & 0.02 \\
\hline
B0528& 0.6.3.7     &  0.23 & 0.36 & 0.07 & 0.05 \\
          & 8.9,11.7    &       & 0.33 & 0.69 & 0.46 \\
B2200& -3.8,2.8 & 0.39 & 0.63 & 0.61 & 0.50 \\
\hline
B0212& -14.1,-7.9 & 0.30 & 0.80 & 0.26 & 0.11 \\
          & -1.4,1.0 & & 0.36 & 0.03 & 0.01 \\
          & 1.3,4.9 & & 0.66 & 0.34  & 0.19 \\  
B0224& -17.1,-11.9 & 0.43 & 0.76 & 0.14 & 0.03 \\
          & -11.6,-9.2 &  & 0.60 & 0.33 & 0.20 \\
          & -9.2,-5.1  &  & 0.56 & 0.39 & 0.20 \\
          & -4.9,-2.0 &   & 0.47 & 0.47 & 0.33 \\
          & -2.0,-0.2 &   &0.39 & 0.05 & 0.28 \\
          &  -0.2 .. 2.0 &   &0.42 & 0.09 & 0.03 \\
B0355& -19.2,-15.9 & 0.35 &0.63 &0.18 & 0.07 \\
          & -15.8,-11.7 &  & 0.92 & 0.43 & 0.28 \\
          & -11.1,-9.8  &  &0.59 & 0.34 & 0.18 \\
          & -9.6,-7.3   &   & 0.62 & 0.34 & 0.10 \\
          & -6.0,-1.4   &  & 0.79 & 0.17 & 0.04 \\
\hline
\end{tabular}}
\\
$^a$ fraction of mapped area with \WCO\ $\geq 1$ \Kkms \\
$^b$ fraction of mapped area with \WCO\ $\geq 2$ \Kkms \\
\end{table}

\subsection{Molecular gas properties in the current sample}

The sightlines studied here were selected on the basis of their known
\hcop\ absorption spectra, creating the possibility that the sample is
biased to large molecular fractions and/or strong CO emission.  However, it
was earlier noticed in a flux-limited survey \citep{LucLis96} not based on
prior knowledge of CO emission that very nearly all sightlines at galactic
latitudes within about 15\degr\ of the galactic equator show \hcop\ 
absorption.  Our present tally, slightly extending the earlier result, is
that \hcop\ absorption occurs toward 19 of 19 sources at $|b| \la 12$\degr,
toward 22 of 25 sources at $|b| \la 18$\degr\ and toward 4 out of 12
sources at $|b| \ga 23$\degr\ including three shown here.  Thus it is a
near certainty that \hcop\ absorption would be detected over the entirety
of the sky fields mapped here below about 15\degr{}-18\degr, no matter what
is the covering factor of detectable CO emission.  This is discussed in
Sect. 8 immediately following the more descriptive portions of the text.

If we discuss the mean properties of the ten sightlines in Table 1 having
reliably determined \EBV\ (all except B0355+508 that lies too near the
galactic plane) in the same terms that we used earlier to derive the mean
CO-\HH\ conversion factor in diffuse gas, \citep{LisPet+10}, we derive an
ensemble average

\begin{eqnarray*}
  \frac{N(\HH)}{\WCO} 
    &=& \frac{5.8\times 10^{21}\pcc<\EBV>-<N(\mbox{H I})>}{<2\WCO>} \\
    &=& 1.52 \times 10^{20}\pcc(\Kkms)^{-1},
\end{eqnarray*}
i.e., 25\% smaller than the previous result found a larger sample. In the
same terms, the mean atomic gas fraction is $<$N(H I)$>$/ $<$N(H)$> =
0.74$, as opposed to 0.65 found earlier.

Estimates of N(\HH) based on assuming the ensemble-average mean value
\citep{LisPet+10} N(\hcop)/N(\HH) $ = 3\times 10^{-9}$ along each line of
sight are also given in Table 1.  They indicate higher molecular fractions
and somewhat higher total column densities N(H) than are found using scaled
\EBV\ for N(H) and using the decomposition discussed just above based on
subtracting N(H I) from N(H) determined as the scaled \EBV.  Specifically
the chemistry-based ensemble average is $<$\fH2 $>$ = $<$2N(\HH)$>$/
$<$(N(H I)+2N(\HH)$>$ = 0.43.

\section{Six simple fields at moderate-high latitude}

\subsection{B0736+0117 $(b \sim 11\degr)$}

The 15\arcmin\ sky field around B0736+016 shown in Fig. 2 is the smallest
and kinematically simplest field studied; the map was made on the spur of
the moment in a relatively brief open period between two other larger maps.
The reddening is modest over the area of the CO map shown in Fig. 2, \EBV
$\la$ 0.165 mag but the molecular fraction implied by the entries in Table
1 is of order 30-50\%.  Toward the source the integrated CO is fairly weak,
$< 0.8$ \Kkms{} but a very slightly blue-shifted 4.5 K line is found within
just a few arcminutes.

\subsection{B0954+658 $(b \sim 43\degr)$}

\begin{figure*}
  \includegraphics[height=13cm]{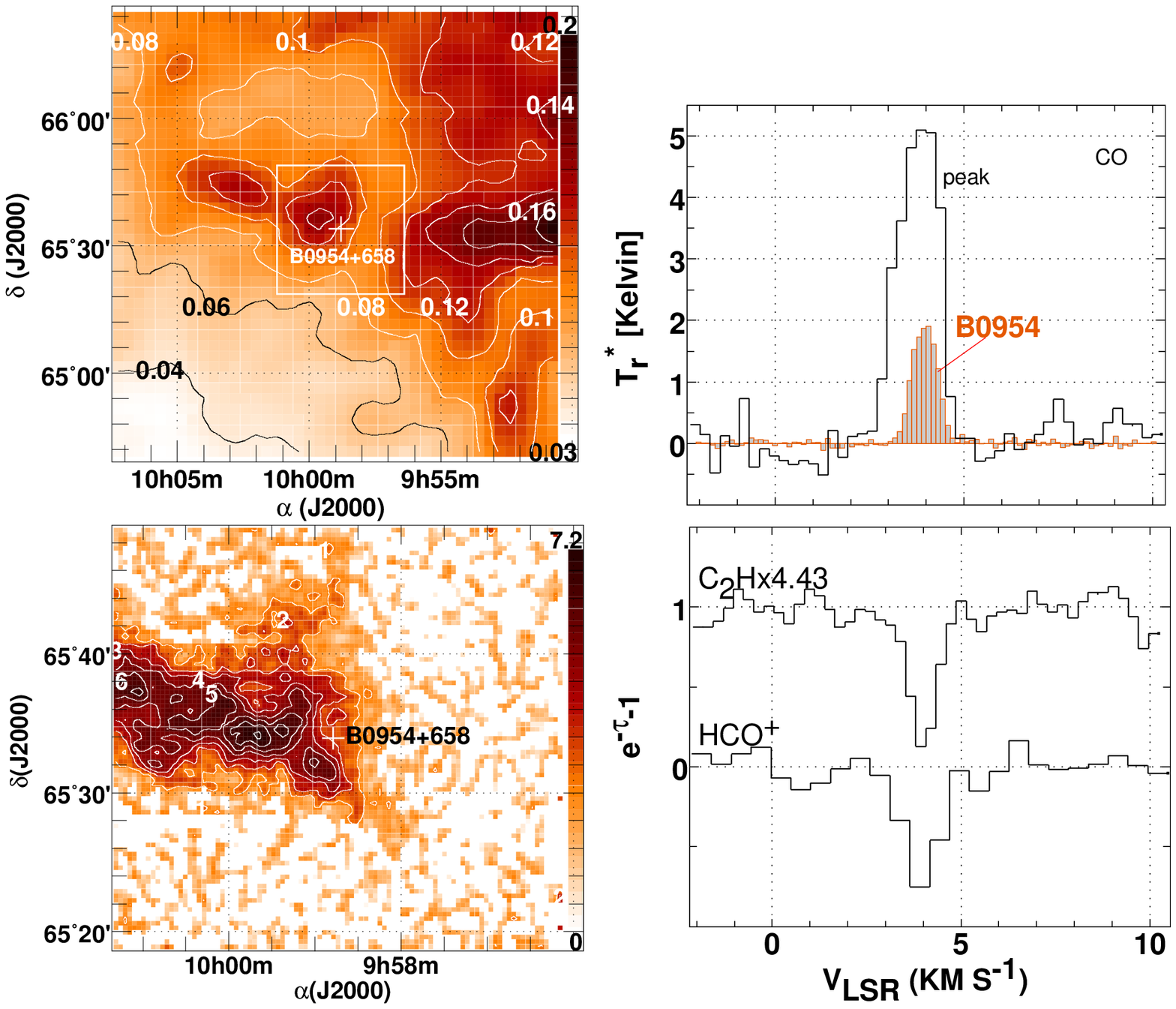}
\caption[] {The sky field around the position of B0954+658, as
  in Fig. 2.}
\end{figure*}

\begin{figure*}
  \includegraphics[height=13cm]{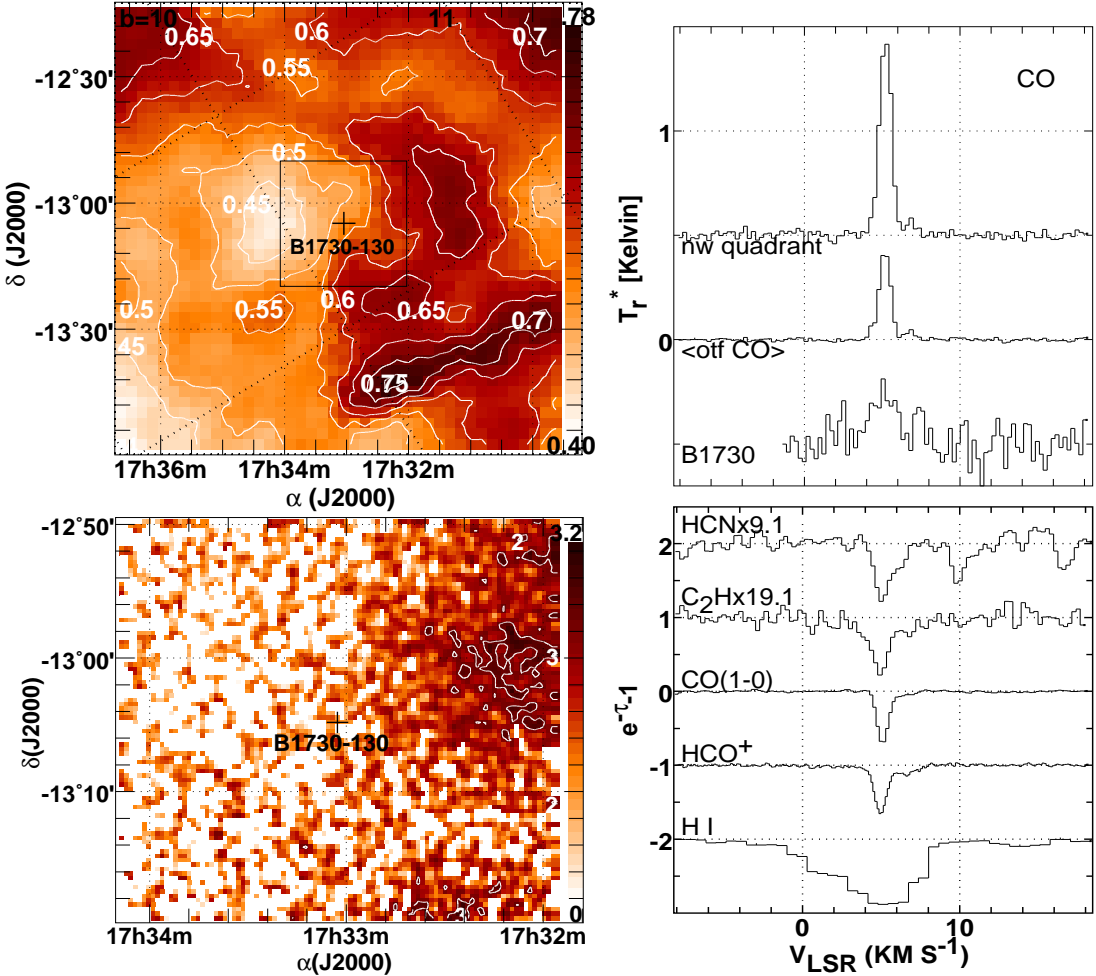}
\caption[]{The sky field around the position of B1730-130, as
  in Fig. 2.  The emission profile labelled ``nw quadrant'' is an average
  over that portion of the map.  The emission profile labelled $<$otf {\bf
    CO}$>$ is the mean over the entire region mapped in CO. }
\end{figure*}

\begin{figure*}
  \includegraphics[height=13cm]{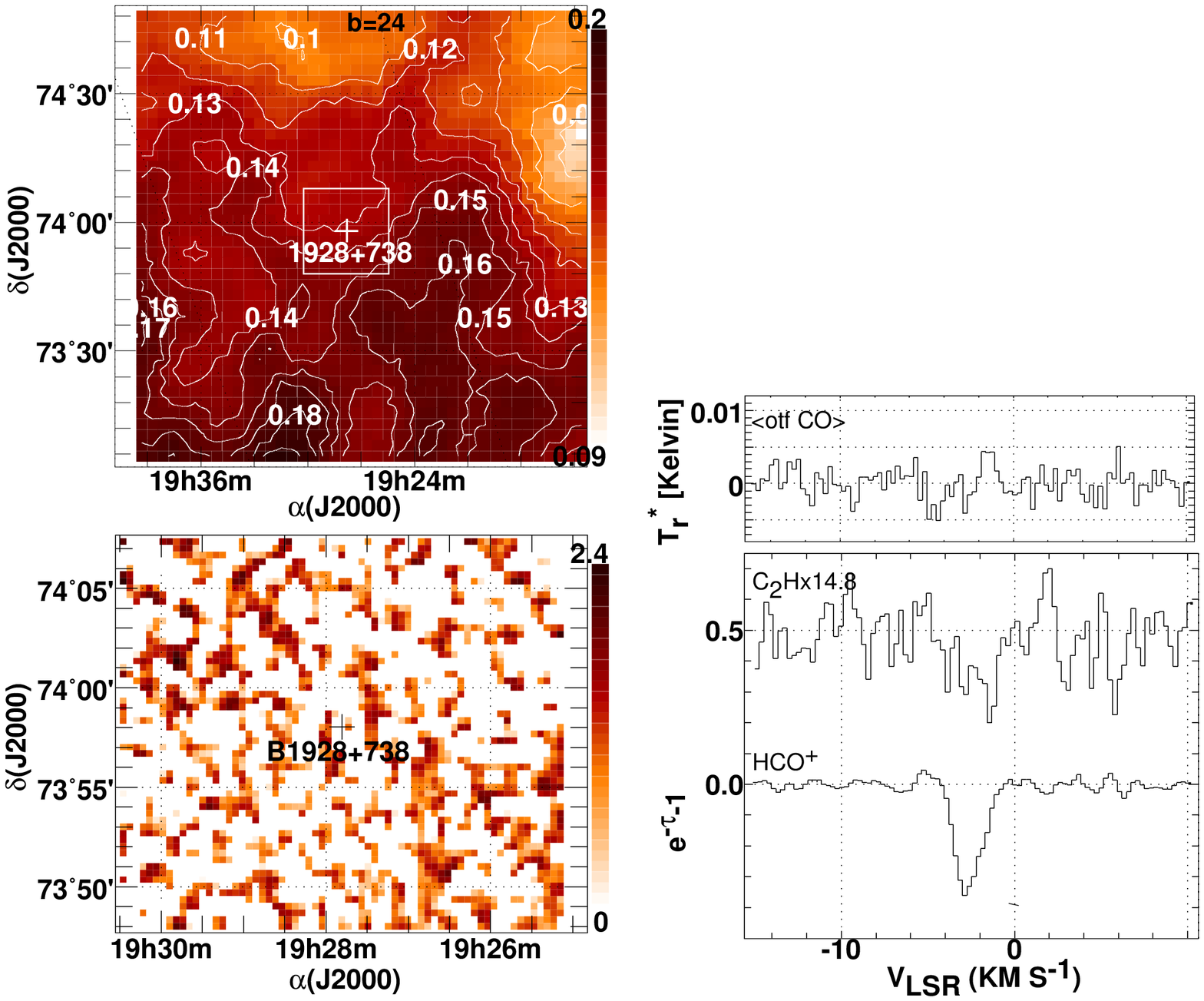}
\caption[]{The sky field around B1928+738. The emission profile at 
  upper right labeled $<$otf {\bf CO}$>$ is the mean over the 20\arcmin
  $\times$ 20\arcmin region mapped in CO. }
\end{figure*}

This source (Fig. 3) is seen at the upper tip of the Polaris Flare near the
location of the M81 group and the so-called small area molecular structures
mapped by \cite{Hei04} (see the finding chart in Fig. 1).  As toward
B0736+017, a fairly strong CO line, 5 K, is seen within a few arcminutes of
the background target where the CO brightness is more modest, 2 K.

The reddening is very moderate but a high molecular fraction, above 50\%,
is suggested by comparing the value of N(\HH) in Table 1 with N(H I) or
\EBV. Consistent with this high molecular fraction and the relative
simplicity of a higher-latitude line of sight, this field is the only one
studied in which there is a strong proportionality between \EBV\ and \WCO,
as discussed in Sect. 8.  The complement of supporting material is
disappointingly slender.

\subsection{B1730-130 $(b \sim 11\degr)$}

The reddening is relatively large over this field (Fig. 4) and the \hcop\ 
absorption is strong, implying a molecular fraction of order one-third, but
CO emission is very weak toward the background continuum source (\WCO\ =
0.4 \Kkms) and absent over most of the field mapped.  Much stronger but
still rather weak emission (1.5 K) is seen 15\arcmin\ to the northwest as
indicated in Fig.  4.  All of the absorption profiles shown in Fig. 4 have
a redward wing that is separately visible in the two spatially-averaged CO
profiles shown at upper right.

\subsection{B1928+738 $(b \sim 23.5\degr)$}

CO emission is absent over the entire 20\arcmin\ field shown in Fig. 5
despite the presence of fairly strong \hcop\ absorption and a suggested
molecular fraction approaching 40\%.  The reddening is modest,
approximately 0.15 mag around the target, but the implied CO-\HH\ 
conversion factor is very large; from the entries in Table 1 we have
N(\HH)/\WCO\ $> 2.5\times\ 10^{21}\pcc$ at the $2\sigma$ level.

\subsection{B1954+513 $(b \sim 12\degr)$}

The reddening in the field around this source is modest, 0.18 mag (see Fig.
6 and 17), and CO emission toward the background continuum source is
unimpressive (2 K) but the \hcop\ absorption is strong and the molecular
fraction is of order 40\%.  Two kinematic components are found in the field
with 4.5 K peak brightness, only one of which is seen toward B1954 in
either emission or absorption.

\subsection{B2251+158, aka 3C454.3  $(b \sim -38\degr)$}

As shown in Fig. 1, this very strong continuum source is seen about 3\degr\ 
removed from an elongated complex of high-latitude clouds that includes the
objects MBM 53-55 \citep{MagBli+85} and new clouds discovered by
\cite{YamOni+03}.  B2251+158 lies within the region surveyed by
\cite{YamOni+03} in CO but the emission detected here (see Fig. 7) escaped
their notice, presumably because of their 4\arcmin\ map sampling of the
2.7\arcmin\ beam.  The reddening is moderate, \EBV\ $\la$ 0.11 mag (see
Fig. 7) and CO emission toward the continuum source is quite weak (0.8 K).
However, much stronger emission (5 K) is seen only 5\arcmin\ away, as with
B0736, B0954 and B1954.  There is no obvious large-scale correlation of the
CO emission with reddening, as evidenced by the weakness of the CO at the
position of highest reddening in the larger field shown at upper left in
Fig. 7 (i.e. the spectrum labelled ``NW'' at upper right there).  The
relationship between \WCO\ and \EBV\ is shown in Figs. 16-17.

The blue wing of the peak emission and the line seen at the northwest
reddening peak both fall to the blue of the CO emission or absorption seen
toward B2251. Nonetheless they overlap a weaker blue wing of the \hcop\ 
absorption that has no counterpart in CO emission, and they fill in a
portion of the H I absorption spectrum.

\section{Two unusual fields at moderate latitude}

\subsection{B0528+134 $(b \sim -11\degr)$}

Mm-wave absorption toward B0528+134 (Fig. 8) was first discussed by
\cite{HogDeG+95}.  This object is viewed against the outer edge of the dark
cloud B30 in the $\lambda$ Orionis ring of molecular clouds
\citep{MadMor87} that is centered on the H II region S264 and its central
ionizing star Lambda Ori (Fig. 1). There is a very substantial foreground
reddening \EBV\ = 0.86 mag and much more heavily extincted regions in the
field to the South.

Although CO emission toward B0528+134 is fairly weak, 2.3 K, emission over
the surrounding field is characterized by a pronounced quasi-periodic
pattern with some very strong (10-12K) and narrow CO emission lines:
emission is undetectable over much of the intervening troughs.  A similar
wavelike pattern may have been observed across the surface of the Orion
molecular cloud by \cite{BerMar+10}.

A weak blue-shifted component of \hcop\ absorption that is absent in CO
toward B0528 has a very bright CO emission counterpart to the Southeast as
shown in Fig. 8.  Despite an 8 \kms\ velocity difference, the blueshifted
emision line gives the strong visual impression of being physically
associated with the main kinematic component at 10 \kms, see the map at
lower left in Fig.  8.  The kinematic span of the CO emission seen at top
right in Fig. 8 neatly coincides with the extent of the H I absorption
toward B0528+134.  Line kinematics in this field are illustrated in more
detail in Fig. 14.

\subsection{B2200+420 aka \bll{} $(b \sim -10.5\degr)$}

This target (see Fig. 9) was the first source seen in mm-wave absorption
from diffuse gas \citep{MarBan+91}, in CO actually, and was also the first
seen in \hcop\ absorption in our work \citep{LucLis93}.  CO emission toward
the source is fairly strong, 4 K or 6 \Kkms{} and the line is quite opaque.
The molecular column density indicated by the strong \hcop\ absorption is
about as large as N(H) inferred from \EBV\ = 0.32 mag, given the \EBV-N(H)
relationship N(H) $= 5.8\times 10^{21}\pcc\,\EBV$ of \cite{SavDra+77}.

The CO emission in this field originates from an unusual filamentary
morphology (Fig. 9 at lower left) at the edge of an arched pattern in the
reddening map.  The integrated intensity takes on very large values within
the field, up to 20 \Kkms\ but the profile is compound and relatively
broad.  Toward the continuum source only the blue side of the core of H I
absorption is seen strongly in molecular absorption or CO emission but a
red-shifted CO emission component overlaying the red side of the H I line
core is present to the Northeast as indicated in Fig 9.

\section{Three complex fields at low-moderate  latitude}

\begin{figure*}
  \includegraphics[height=15cm]{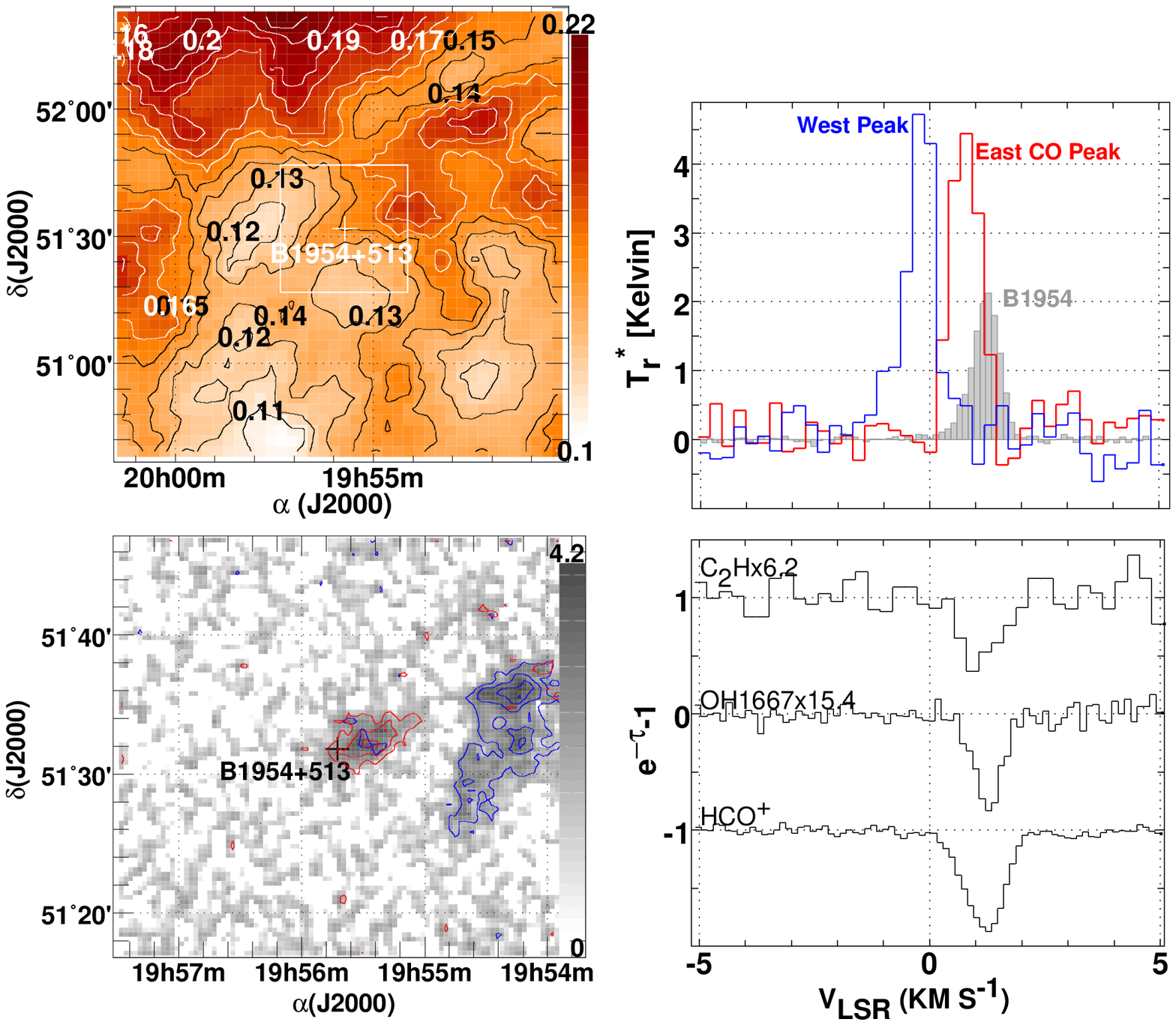}
\caption[]{The sky field around the position of B1954+513, as in Fig. 2.  
  In the map at lower left the grayscale represents the total integrated
  emission at -1$\le$ v $\le$ 2 \kms\ and the red and blue contours show
  the individual distributions of red and blue-shifted components,
  respectively.  Profiles at the peak of the red and blue-shifted emission
  components are shown at upper right along with the profile toward the
  continuum source (shaded).}
\end{figure*}
 
\begin{figure*}
  \includegraphics[height=15cm]{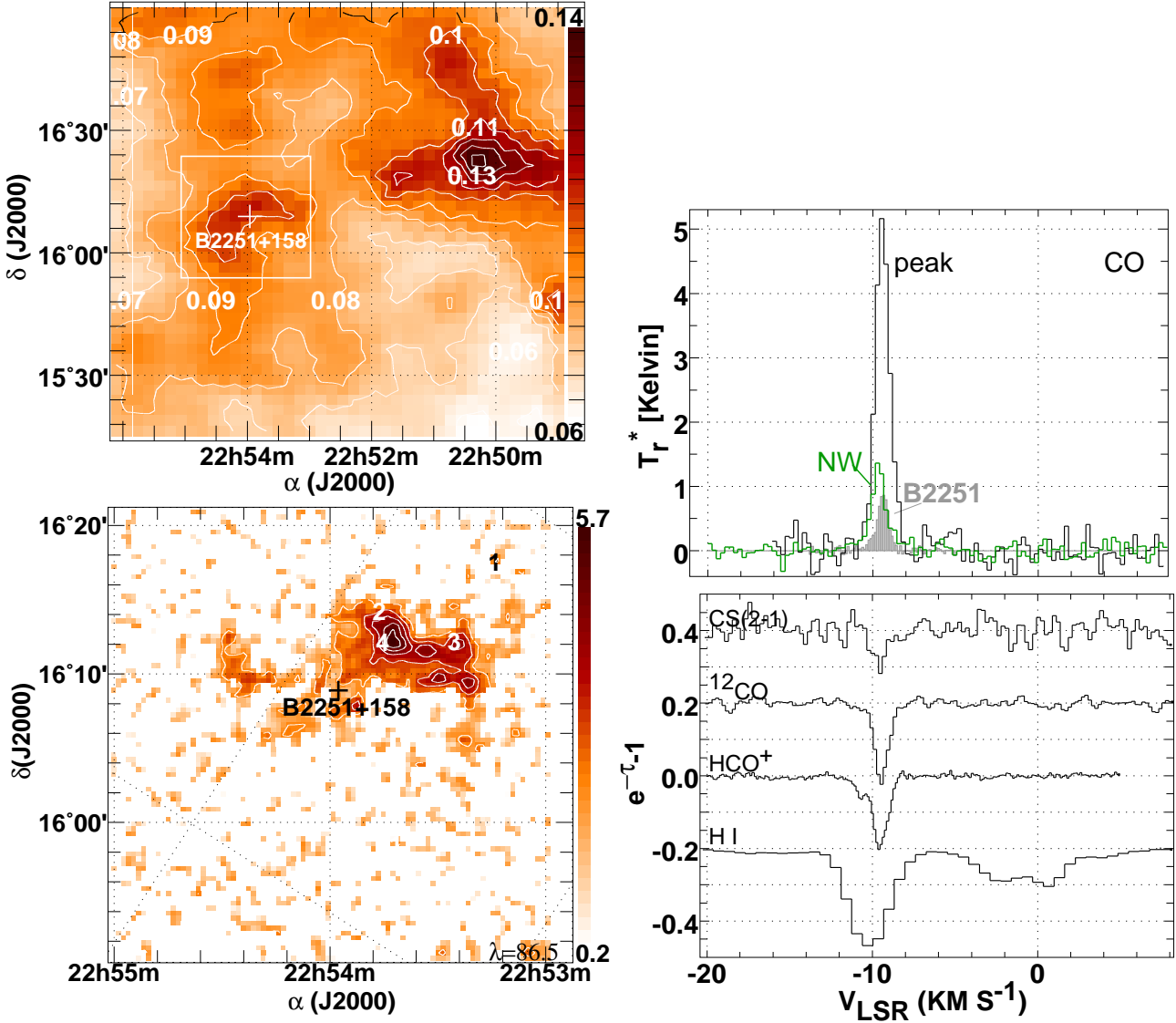}
\caption[]{The sky field around the position of B2251+158 (aka 3C454.3), as in 
  Fig. 2.  The map of reddening at upper left is offset to show a separate
  peak to the Northwest near $22^{\rm H}50^{\rm M}$ and a profile at the
  position of this peak is shown at upper right, shaded green and labeled
  'NW', along with profiles toward 3C454.3 (shaded) and at the peak of the
  small clump that is seen immediately adjacent to the continuum source.}
\end{figure*}

\begin{figure*}
  \includegraphics[height=15cm]{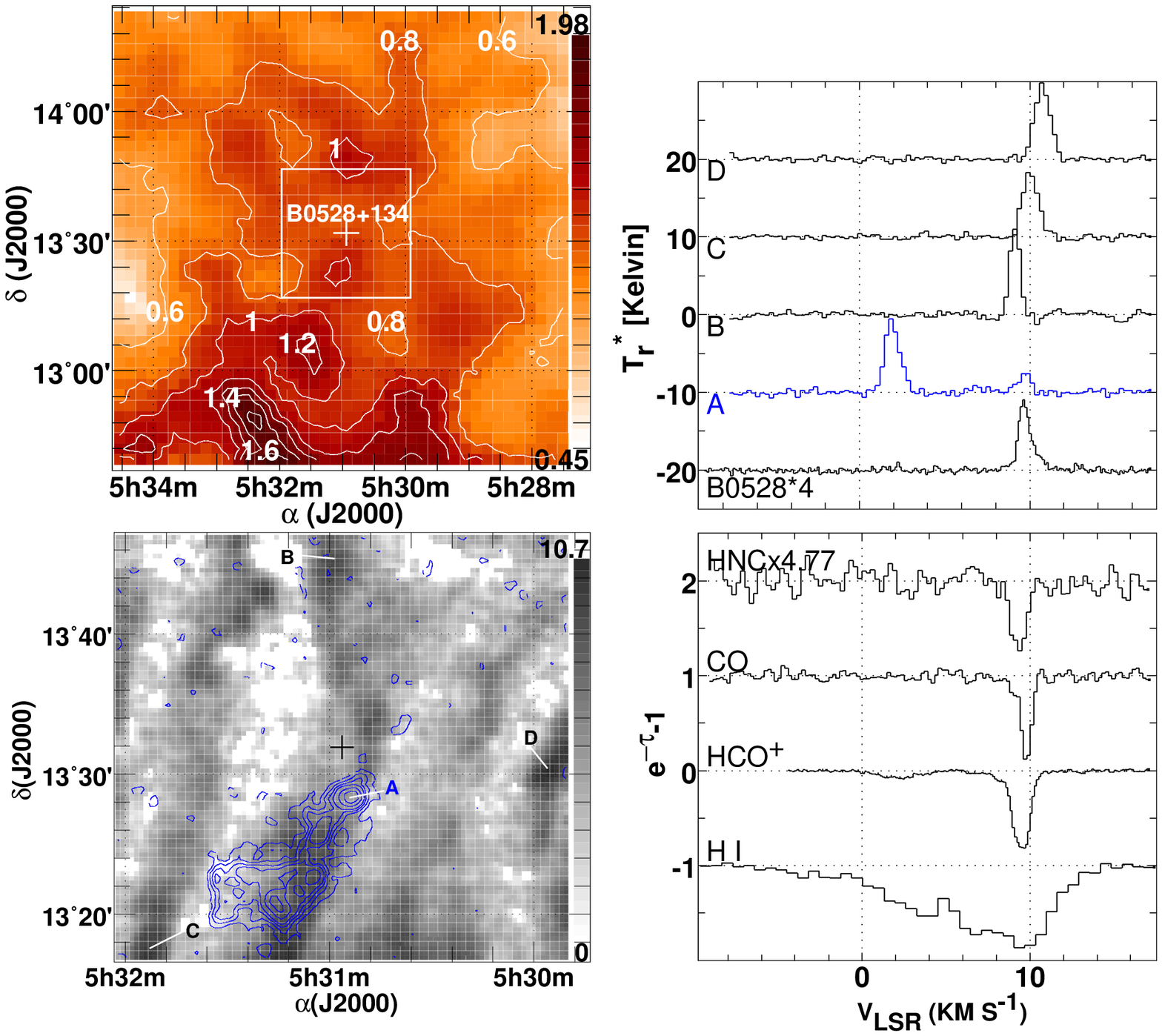}
\caption[]{The sky field around the position of B0528+134, 
  as in Fig. 2.  The map of CO emission at lower left superposes the
  integrated intensity at 0-4 \kms\ as blue contours against a background
  grayscale representing emission at v = 8-12 \kms.  Very strong CO lines
  are seen in the foreground gas as shown in the upper right panel:
  positions at which they originate are indicated at lower left.}
\end{figure*}

\subsection{B0212+735 $(b \sim 12\degr)$}

B0212+735 (Fig. 10) sits in a mild trough with \EBV\ $\approx$ 0.76 mag in
a region of substantial reddening at moderate galactic latitude b=12\degr.
It has three molecular absorption components whose balance is entirely
opposite to that of H I.  Whereas most of the atomic absorption toward
B0212+725 occurs in a deep and broad feature at v $\la 10$ \kms, most of
the molecules are concentrated in a narrower-lined feature at v $\approx$ 4
\kms.  An obvious molecular absorption feature at 0-velocity is, very
unusually, not apparent in H I.  It seems possible that the low velocity
resolution of the H I profile (1 \kms) is responsible. The only other
published example of this phenomenon is toward B0727-115 \citep{LeqAll+93}.

The CO emission line kinematics have been color coded at lower left in Fig.
10 to display the observed behaviour in one panel.  The gray-scale
background represents the integrated intensity of the gas at 1.5-5 \kms;
higher resolution mapping with the IRAM 30m telescope to be discussed in a
forthcoming paper indicates that the feature is compound but this is not
apparent in the present dataset.  The blue contours represent the CO
profile integral at $-16$ \kms{} $\leq$ v $\leq$ $-9.5$ \kms; consistent
with the prominence of this gas in H I, it is almost as widely distributed
over the field as the stronger emission at 1.5-5 \kms\ (Table 2: 26\% vs.
34\%) even if it is barely seen toward the continuum.  The profile labelled
'A' at upper right is an example.  The black contours represent the profile
integral at $-2$ \kms\ $\leq$ v $\leq$ 1 \kms{} and an example is shown at
upper right as profile 'B'.  Emission from this gas occurs only at the
eastern edge of the map area.

\begin{figure*}
  \includegraphics[height=15cm]{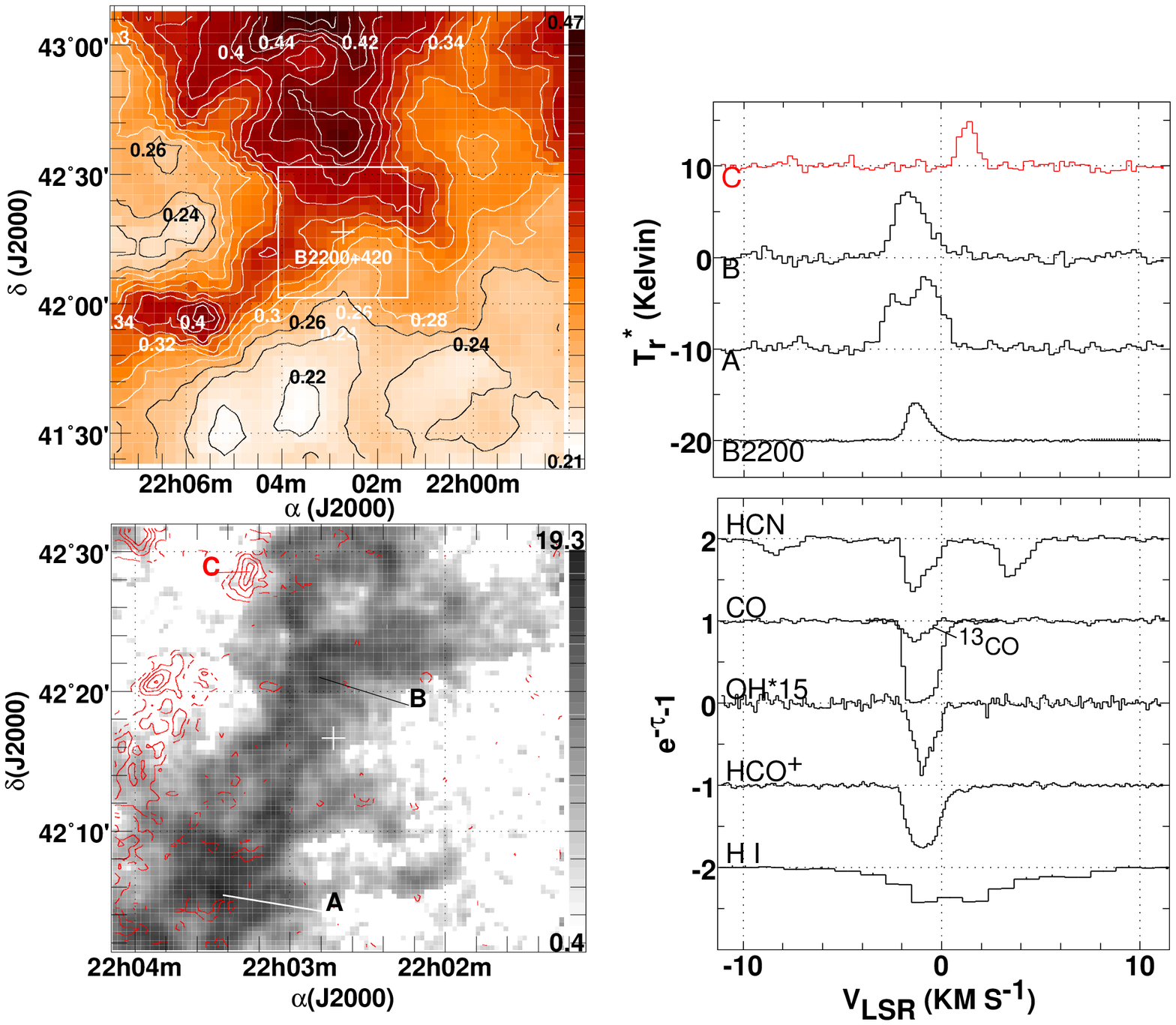}
\caption[]{The sky field around the position of B2200+420 (aka \bll), 
  as in Fig. 2. The map of CO emission at lower left superposes the
  integrated intensity in the range v=0-2 \kms\ as red contours against a
  background grayscale representing emission at v $\leq 0$ \kms. Molecular
  absorption and most emission is sequestered in the blue wing of the core
  of the HI absorption profile but a red-shifted emission component is
  present to the Northeast as illustrated by the spectrum at position ``C''
  indicated at lower left.}
\end{figure*}

\begin{figure*}
  \includegraphics[height=15cm]{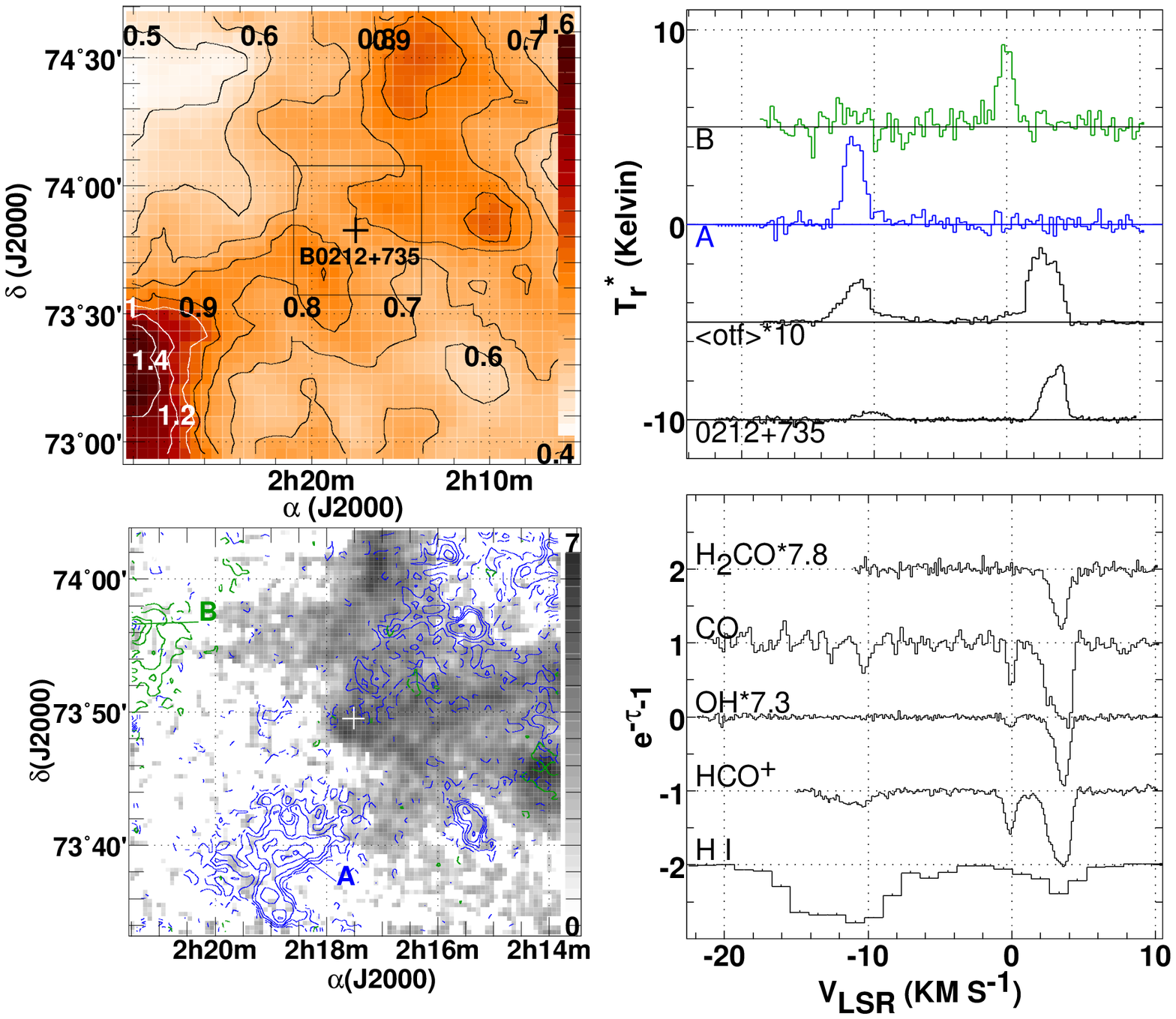}
\caption[]{The sky field around the position of B0212+735,
  as in Fig. 2.  Contours in the CO emission map at lower left are color
  coded in blue for emission at $-15 \leq {\rm v} \leq -9.5$ \kms\ and in
  green for emission at $-2 \leq {\rm v} \leq 1$ \kms.  The gray scale
  background represents the integrated emission of the strongest emission
  component seen toward the continuum source, at v = 1.5 - 5 \kms.  \cotw\ 
  spectra at two locations labelled A and B are shown at upper right along
  with a strongly-scaled mean profile taken over the full map area.}
\end{figure*}

\subsection{B0224+671 $(b \sim 6\degr)$}

This line of sight toward B0224+671 (Fig. 11) samples the two
lower-velocity features seen toward B0212+735 but at substantially lower
galactic latitude b=6.2\degr, see Table 1.  The extinction is large in this
field as are the H I and inferred \HH\ column densities.  CO emission is
weak on a per-component basis toward the continuum target but fairly total
large values of \WCO\ are attained overall.

The integrated CO emission is compact but rather formless because it is the
sum of many kinematic components.  Paradoxically, the strongest molecular
features seen toward and near B0224+671 are not widely distributed over the
map area as shown in the middle panel at right in Fig. 11 comparing the
profile toward B0224+671 with the unweighted average of all profiles
denoted '$<$otf$>$': the strongest emission peak, at the red edge of the CO
emission profile toward B0212+735 is strongly underrepresented in the mean
profile.  Examples of profiles seen over the map area are shown at upper
right in Fig.  11; they were chosen at local peaks in more finely-divided
(in velocity) moment maps, with velocity increasing from a to g.
Especially at v $<$ 0 \kms\ the lines shown are much stronger than seen
toward the continuum.  Among them, the various CO emission components cover
the range of strong H I absorption toward the continuum source, $-15$ \kms\ 
$\leq$ v $\leq$ 2 \kms{}. B0224 and B0212 have similar absorption spectra
in that both have stronger atomic absorption at v $ < -10$ \kms{} where the
molecular absorption is weaker. They thus sample the same large-scale gas
kinematics although they are separated by about 15 pc, assuming they lie on
a sphere of 150 pc radius.

\subsection{B0355+508 aka NRAO150 $(b \sim -1.6\degr)$}

This is the only low-latitude source studied here.  The more strongly
blue-shifted gas seen in this direction is likely to be relatively distant.
The actual velocity field is probably affected by galactic streaming
motions but a typical velocity gradient due to galactic rotation in this
direction is 8\kms\kpc$^{-1}$.

\begin{figure*}
  \includegraphics[height=15cm]{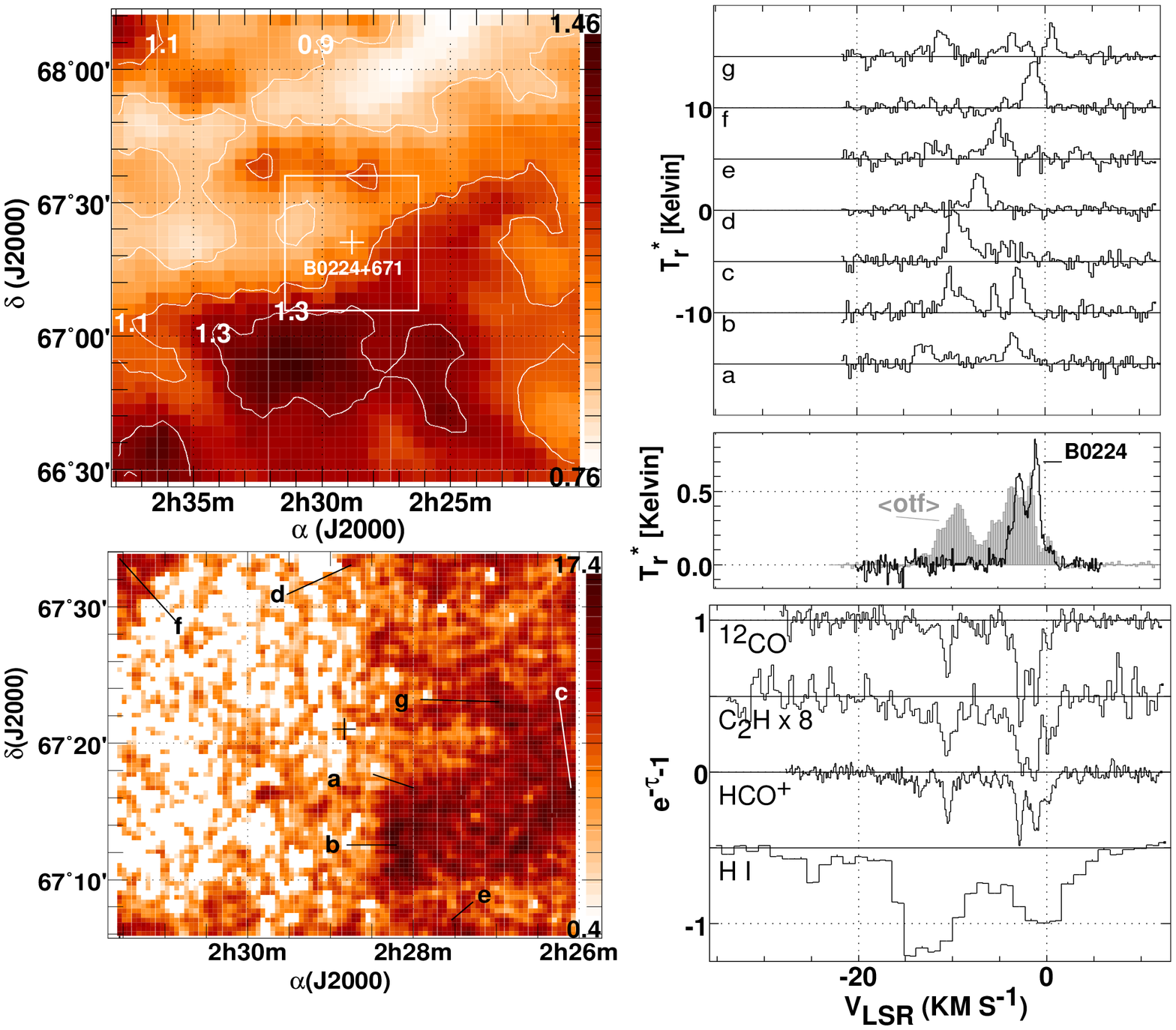}
\caption[]{The sky field around the position of B0224+671,
  much as in Fig. 2. The map at lower left has been integrated over the
  very wide interval $-15.5 \leq v \leq +2$ \kms.  Shown in the middle
  panel at right are the CO emission spectrum toward B0224+671 and as
  averaged over the region of the entire CO emission map.  At top right are
  example profiles from the positions labelled at lower left, chosen from
  moment maps over narrow intervals increasing in velocity from a-g.}
\end{figure*}

\begin{figure*}
  \includegraphics[height=13.3cm]{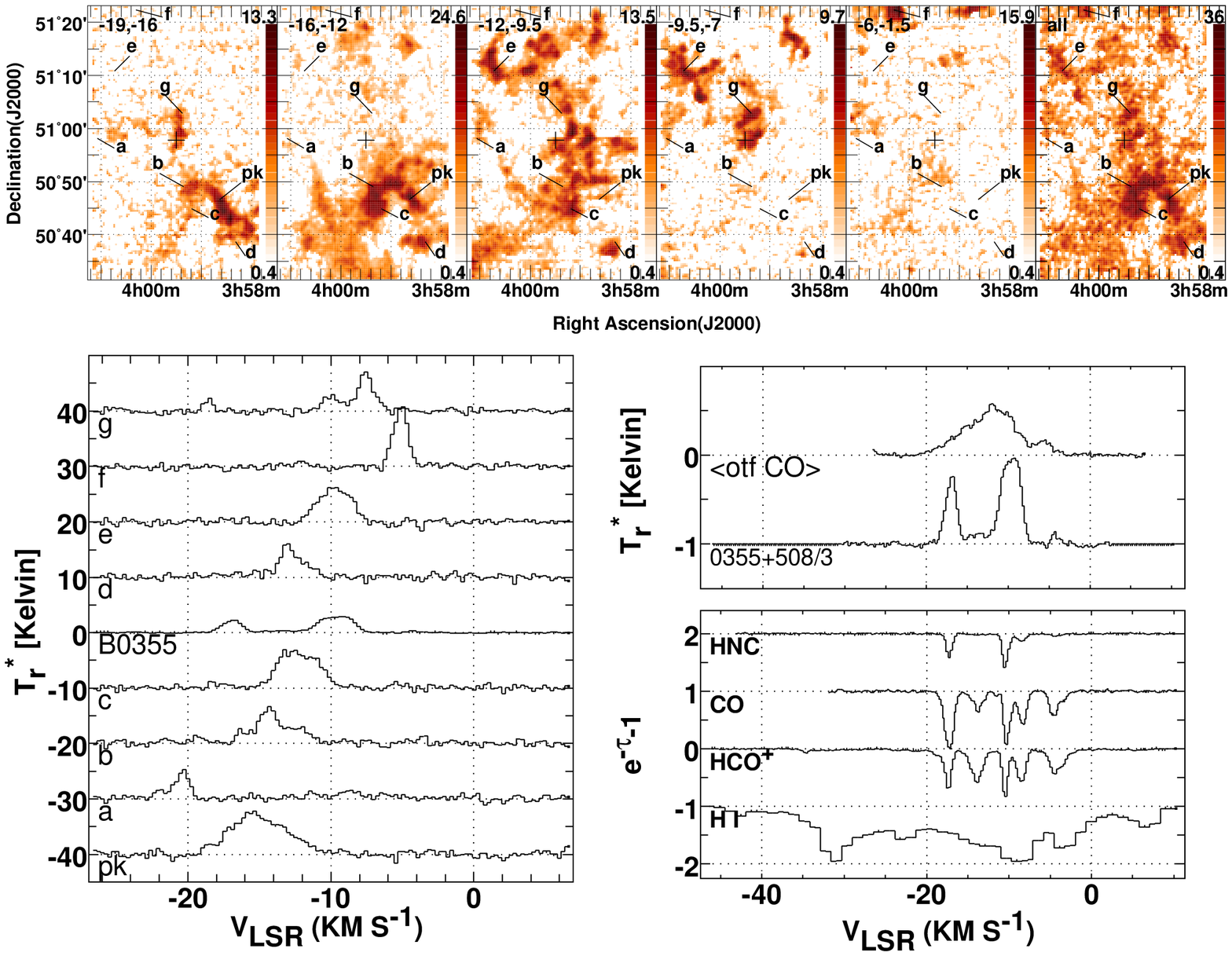}
\caption[]{The sky field around the position of B0355+508. At top 
  are maps of integrated CO intensity made over the velocity intervals
  indicated in each panel, corresponding to the five strong components of
  the \hcop\ absorption profile seen at lower right.  CO emission profiles
  at various locations indicated in the map panels are shown at lower left.
  CO emission profiles toward B0355+508 and averaged over the map area are
  shown above the absorption line profiles.}
\end{figure*}

Fig. 12 does not display a map of the reddening for this source because the
reddening maps are not believed to be accurate at such low galactic
latitudes.  Taken as a whole the line of sight is very heavily extincted,
with large columns of both H I and \HH\ (Table 1).  However, the high
inferred \HH\ column density is the sum of components whose individual
values indicate that they have \AV\ $\approx 1$ mag.

The absorption profiles toward B0355+508 at lower right in Fig. 12 show
five obvious \hcop\ and CO components having roughly equal N(\hcop)
$\approx 1.2\times10^{12}\pcc$ and somewhat more variable N(CO)
\citep{LisLuc98}.  Only two have substantial abundances of less-common
species such as HCN.  Less obvious is the fact that the \hcop\ absorption
profile has a weak broad blue wing extending to $-35$ \kms{} so that the
entire core of the H I absorption line is seen in molecular gas
\citep[see][]{LisLuc00}.

Shown at the top in Fig. 12 are CO moment maps made over velocity intervals
corresponding to the obvious \hcop\ and CO absorption line features; a map
integrated over all velocities is shown in the top right-most panel. The CO
emission distribution is heavily structured and very complex.  Profiles at
positions of local peaks in narrower CO moment maps are shown at lower
left, along with profiles at the integrated emission peak (see Fig. 12 at
top right) and toward the background source.  At the eastern edge of the
map at position ``a'' there is a blue-shifted CO emission line that
(unusually) falls outside the range of the strong molecular absorption
toward B0355+508 but is well inside the H I absorption profile).  A very
bright ($>$ 10 K) line is found at v = $-4.5$ \kms\ at the northern edge of
the map at position f, corresponding to a prominent molecular absorption
feature that has only a very weak emission counterpart toward the
continuum.
 
The middle right panel displays the profile toward the continuum background
source and the profile averaged over the mapped field of view. The
kinematics of this region are shown in more detail in Fig. 15 and discussed
in Sect. 7.2.  The displacement of the two strong CO emission lines about
the centroid of the mean profile results from a coherent kinematic
patttern, perhaps a shell or bubble in the underlying gas distribution.
Recall, however that absorption at the mean field velocity is not absent
toward the continuum source.

The complexity of the emission distribution makes the division into ranges
based on \hcop\ absorption quite arbitrary.  Moreover, the emission and
absorption profiles show rather different structure even toward the
continuum target.  A very detailed discussion of CO emission within a
90\arcsec\ field centered on NRAO150 was given by \cite{PetLuc+08}.
Remarkably, the peak emission brightness seen just $6''$ from the
background continuum source is almost 13 K.  As the spatial resolution
increases, the CO emission profile toward B0355 more nearly resembles the
absorption and the blended emission at v $\approx$ -10 \kms\ resolves into
two distinct components.

\begin{figure}
  \includegraphics[height=7.2cm]{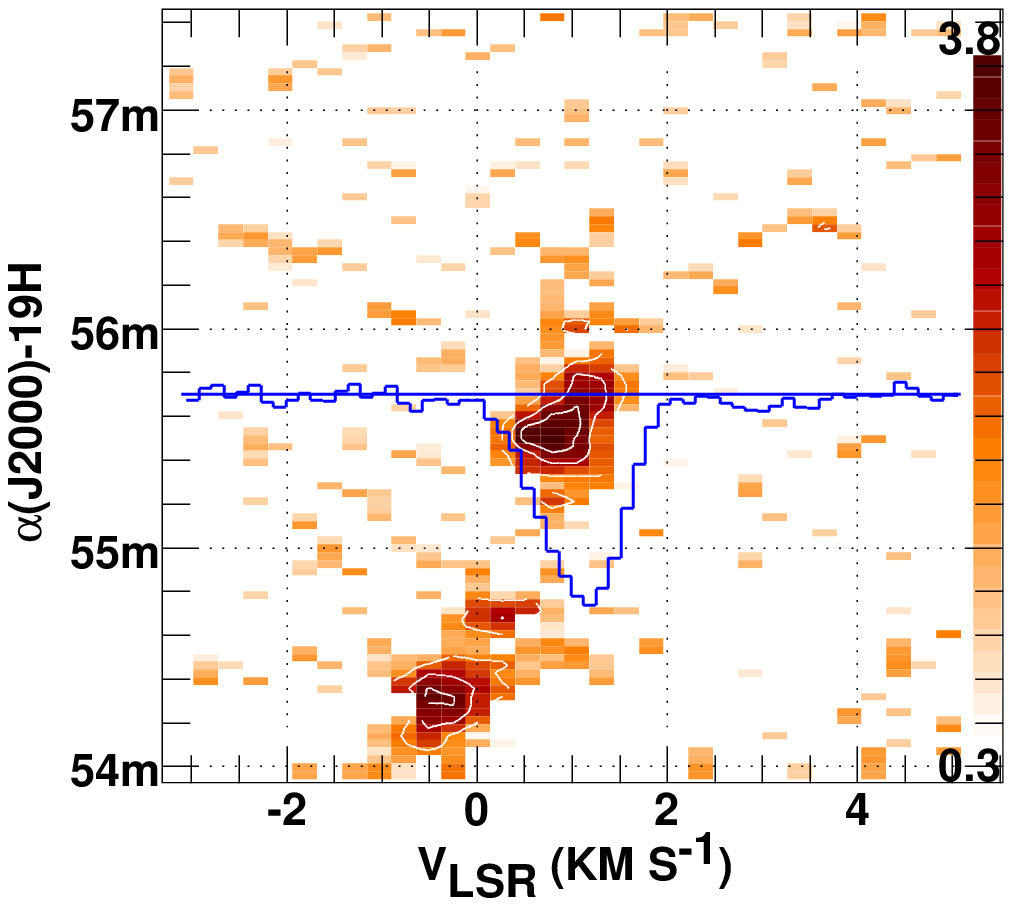}
\caption[]{A right ascension-velocity diagram of CO emission through 
  the position of B1954+513.  The \hcop\ (not CO) absorption spectrum
  toward B1954 is shown with its 0-level at the location of the continuum
  source; the peak absorption is 90\%.  Contours are shown at levels 1, 2,
  3, ... K.}
\end{figure}

Conversely, in the present dataset, the emission components seen toward
B0355+508 lose their identity as the resolution degrades below a 4\arcmin\ 
hpbw.  Both the peak profile and the mean are broad, largely unstructured
and centrally peaked about velocities lying between the two strong CO
emission components seen toward the background.

\section{Statistical lessons}

Faute de mieux, the first surveys for suitable absorption-line targets were
conducted in CO emission \citep{BanMar+91,LisWil93,Lis94} but the discovery
of yet more common \hcop\ absorption \citep{LucLis96} caused a reversal in
the search strategy for diffuse molecular gas.  Thus, all targets studied
here were pre-selected to have absorption from \hcop\ but only some were
known to have CO emission.  However any division between targets with and
without CO emission is misleading because the same sightline may have and
lack CO emission on a per-component basis.

Stronger CO emission is always found somewhere else in the map when CO is
present toward the continuum but comparably strong CO emission was found,
somewhere on the sky, from absorption components lacking CO emission
counterparts toward the continuum background.  This is true with only one
exception, B1928.  The implication is that CO emission is somewhat more
ubiquitous than is presently believed to be the case because nearly all
\hcop\ absorption components will be found in nearby emission after a small
search.  In terms of numbers, in this work we observed 20 absorption line
components (and a few distinct line wings) with 13 carbon monoxide emission
counterparts toward the continuum targets, and we found 23 CO emission
components within $15'$ of the background continuum source during the
mapping.

The following gives some conclusions that are drawn from the preceding
presentation; they should generally be understood as applying on a
component-by-component basis.

{\bf From the standpoint of CO emission:}

1) In every case where CO emission was detected toward the background
source, much stronger emission was also detected within approximately
15\arcmin\ and often much less.

2) Near B0355+508, B0528+134 and B2200+430 the nearby stronger emission was
very strong indeed, with peak line temperatures of 10-12 K and/or line
profile integrals as large as 20 K \kms.

3) 1) and 2) are also generally true for kinematic components that were
present in absorption toward the continuum source but {\it not} detected in
emission there (7 of 20 components).

4) In one case only (B1928+738), representing 1 of 11 fields and 1 of 20
kinematic absorption line components, the entire field mapped was devoid of
emission when emission was not detected toward the continuum source (from a
kinematic absorption line component).  The field mapped around B1928 was
only 20\arcmin\ $\times$ 20\arcmin\ as against 30\arcmin\ $\times$
30\arcmin\ or more for all of the other fields.

5) In 2 fields we found an emission feature without a counterpart in
molecular absorption toward the background continuum object (B1954 and
B0355 at -20 \kms).

{\bf From the standpoint of absorption:}

1) The same kinematic components are seen in both absorption and emission
with angular separations of up to $15'$.

2) The absorption spectra toward a background source are a preview of what
will be seen in emission in a larger field about the background source.

3) Molecular absorption components seen toward the continuum source were
found in CO emission somewhere in the field except in the smaller region
mapped around B1928+738

{\bf From the standpoint of the atomic-molecular transition:}

1) The same kinematic components are seen in both atomic and molecular
tracers at angular separations between $0'$ and $15'$.

2) The components seen in molecular absorption are present in H I
absorption, although somewhat indistinctly in some cases.  For instance,
the 0-velocity molecular absorption line in B0212+735 appears only as a
blue wing of the 4 \kms\ H I absorption component.

3) Portions of H I absorption profiles adjacent to molecular features but
lacking a molecular counterpart are seen in CO emission elsewhere in the
field in two cases on (B2200+420 and B0528+134).

4) We saw no molecular features in absorption or emission outside the span
of the H I absorption (see Appendix B).

\section{Kinematics}

Molecular gas is generally well-mixed with other components of the ISM
\citep{DamTha94,GirBli+94} and does not require exceptional kinematics.
This is apparent in our work from the coincidence of molecular and atomic
absorption features, even if they do not have precisely the same patterns
of line depth.  The kinematics are affected by galactic structure and local
external influences such as shocks, but this only becomes apparent on broad
angular scales.  The targets B0212+735 and B0224+671 (Figs 10-11) are
relatively close to each other and both are most strongly absorbed in H I
around -15 \kms.  The background target B2251+158 (Fig. 7 and Sect. 3.6) is
seen in the outskirts of the MBM53-55 cloud complex, which is part of a
large shell that has been extensively mapped in molecular and atomic gas
\citep{GirBli+94,YamOni+03}.

In individual line profiles and over small scales, the kinematics are often
dominated by the internal structure of individual clouds.  The internal
motions of diffuse molecular gas are now understood to reflect turbulent
gas flows \citep{PetFal03,HilFal09} that are characterized by unsteady
projected velocity fields with strong shears and abrupt reversals of the
velocity gradient.  \cite{SakSun03} show the transition between diffuse and
dense molecular gas at the edge of TMC1 and \cite{LisPet+09} discuss gas
flows in the diffuse cloud occulting \zoph.

In this Section we discuss the kinematics of just two of the fields mapped
here.  Further examples of CO kinematics in individual sky fields are given
in Figs. A.1-A.3 of Appendix A (available online) and the galactic context
for all fields is given in Figs. B.1 and B.2 of online Appendix B, showing
large-scale latitude-velocity cuts in HI from the Leiden-Dwingeloo H I
survey of \cite{HarBur97} with the locations of the continuum background
sources marked in each case.

Fig. 13 shows the kinematics in the relatively simple sky field around
B1954+513 (Sect. 3.5 and Fig. 6) with the spatially-displaced blue and
red-shifted CO emission components that were illustrated in Fig. 6.  The
red-shifted component seen toward the continuum has a partially-resolved
velocity gradient that carries it just to the midpoint of the associated
\hcop\ absorption profile at the continuum position.  It is certain that
the blue-shifted CO emission to the East would have an associated \hcop\ 
absorption at its position but the structure of the redward gas cannot be
traced away from the continuum and, regrettably we do not have an H I
absorption profile that might show both the red and blue-shifted gas in
atomic absorption as toward \bll\ (Fig. 9 and Sect. 4.2).

Fig. 14 shows the more complicated field at low latitude around B0355+508
(Sect. 5.3, Fig. 12) and illustrates how the partition of a line profile
into components, no matter how seemingly obvious, can also be arbitrary and
capricious.  None of the well-defined absorption features has an obvious CO
emission counterpart except perhaps in the immediate vicinity of the
continuum target.  This is not an artifact of taking a cut in declination,
which is actually richer than that in right ascension (see Fig. 12).

Nonetheless, mapping the CO emission does help to clarify interpretation of
the absorption profiles.  For instance, consider gas near $-9$ \kms\ around
the location of B0355 in Fig. 14.  In absorption there are two distinct
kinematic components at $-11$ and $-8$ \kms\ that would usually be
interpreted as unrelated because, aside from their separation in velocity,
they have different patterns of chemical abundances (Fig. 12) However, Fig.
14 shows that the CO emission line has an appreciable velocity gradient
across the position of the continuum source, spanning the two absorption
lines, making it likely that the two absorption components are part of the
same body \footnote{\cite{PetLuc+08} show that the overlapping CO emission
  line is resolved into two kinematic components at 6\arcsec\ resolution
  toward the continuum source.}.  Moreover, the CO mapping suggests that
the components at $-17$ and $-10$ \kms\ may also be part of the same
structure (and separated by a velocity gradient), which was actually
suggested by several coincidences in our earlier high-resolution CO mapping
\citep{PetLuc+08}.  The lines at $-11$ and $-17$ \kms\ are very bright (13
K) at high resolution and have considerable chemical complexity.  There are
also some seemingly correlated spatial intensity variations.  The evidence
for an association is entirely indirect but has a clear precedent in the
kinematics around B0528+134 (Sect. 4.1 and Fig.  8) where a similar
velocity separation occurs between two emission components that are seen
superposed in an unusual wave-like spatial configuration.

\section{The brightness of diffuse cloud CO}

\subsection{\WCO\ relative to \EBV\ and \fH2}

The large-scale finding chart in Fig. 1 is a map of the total intervening
gas column density, except where discrete sources of infrared emission
(often H II regions) ``leak'' into the map (usefully indicating when the
background target may have been observed through disturbed foreground gas).
Large-scale surveys of CO emission at 8\arcmin\ resolution show a good
correlation with reddening \citep{DamHar+01}, contributing to the common
interpretation of CO sky maps as displaying the global distribution of
dense, fully-molecular gas.

In diffuse gas appreciable scatter in the \WCO-\EBV\ relationship is
expected because the reddening is a sum over atomic and molecular
components that both make important contributions to N(H), combined with
the fact that both N(\HH) and N(CO)/N(\HH) exhibit order-of-magnitude or
larger scatter with respect to \EBV\ even when all quantities are measured
along the same microscopic sightlines toward nearby bright stars
\citep{BurFra+07,RacSno+09}.  The disparity in angular resolution between
the reddening data and our 1\arcmin\ CO maps presents another sort of
complication that is considered in Sect. 8.2 but does not by itself
dominate the scatter.  Recall also the discussion in \cite{LisPet+10} where
a good correlation was shown between \EBV\ at 6\arcmin\ resolution compared
with the integrated H I optical depth measured in absorption at 21cm toward
a larger set of the same kind of point-like radiocontinuum background
target considered here.

Small-scale maps of reddening are shown in the various Figs. 2-12 detailing
the individual fields.  They may visually suggest correlations between
\EBV\ and \WCO, and there is a threshold \EBV\ $\ga $ 0.09 mag for
detecting CO emission, consistent with the well-known and quite abrupt
increase of N(\HH)/N(H) at comparable reddening \citep{SavDra+77}.
However, reddening is not a reliable predictor of CO emission in our sky
fields.  For instance, in the field around B2251+158 in Fig. 7, CO emission
is much weaker at the peak of the reddening map where \EBV\ = 0.14 mag (the
profile indicated as ``NW'' at upper right in Fig. 7) than nearer the
continuum source at smaller \EBV\ = 0.10 mag.  Around B2200+420 (Fig. 9)
the shape of the CO distribution appears to parallel that of the reddening
but in detail CO only traces the edge rather than the peak ridge of the
\EBV\ distribution.

In Fig. 15 we show the relationship between \WCO\ and \EBV\ in the four
simple cases discussed in Sect. 3, where the extinction is small and a
single narrow CO spectral component is present at each pixel
\footnote{Green diamonds in Figs. 15 and 16 show \EBV\ and \WCO\ toward the
  continuum target as given in Table 1}. The rms noise levels in these four
datasets (Table 2) are 0.48, 0.33, 0.32 and 0.35 \Kkms{} reading clockwise
from upper left so that datapoints with \WCO\ $\ga$ 1 \Kkms\ (the usual
last contour on CO sky maps) are detected at or above the 90\% confidence
level.  To put these brightness and sensitivity levels in context, note
that there is a straightforward relationship between \WCO, \fH2, and \EBV\ 
once the CO-\HH\ and \EBV/N(H) conversion factors are fixed; for the
\emph{standard} \WCO/N(\HH) = $2\times 10^{20} \pcc$ \HH\ (\kms)$^{-1}$ and
N(H)/\EBV\ = $5.8\times 10^{21} \pcc {\rm mag}^{-1}$ one has \WCO\ = 14.5
\fH2\ \EBV \Kkms{}.  At \EBV\ = 0.1 mag, emission only slightly exceeding 1
K \kms\ implies a molecular fraction \fH2\ $>$ 1 and therefore is too
bright to be accomodated by a CO-\HH\ conversion factor as large as the
standard $2\times 10^{20} \pcc$ \HH\ (\kms)$^{-1}$.

\begin{figure}
  \includegraphics[height=8.5cm]{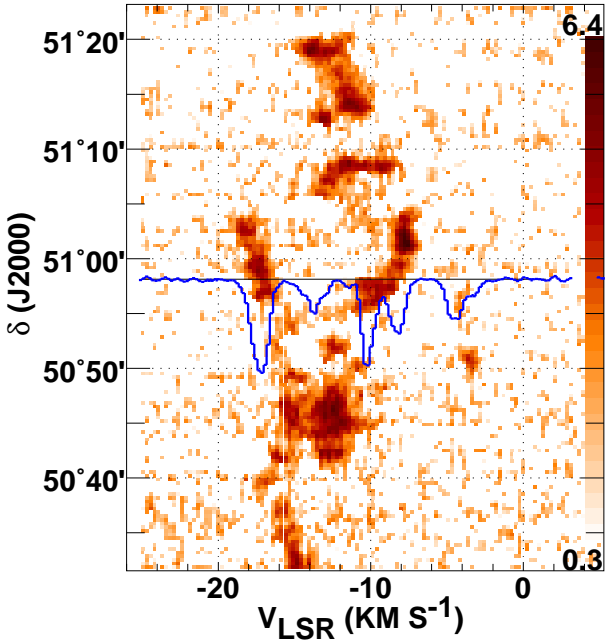}
\caption[]{A declination-velocity diagram of CO emission
  at the right ascension of B0355+508.  The CO absorption spectrum toward
  B0355+508 is shown with its baseline level at the declination of the
  background source.  The strongest CO absorption line is quite opaque, see
  Fig. 12.}
\end{figure}

Shown in each panel of Fig. 15 are lines representing the CO emission
expected if various fractions \fH2\ of the total neutral gas column are in
\HH\ with a typical galactic \WCO/N(\HH) conversion factor X$_{\rm CO} =
2\times10^{20}$ \HH\ $\pcc$ (\Kkms)$^{-1}$.  Much of the CO in Fig. 15
occurs above the line corresponding to \fH2\ = 1 and is therefore too
bright to be accomodated by the usual CO-\HH\ conversion factor; indeed,
almost every CO line with \WCO\ $\ga$ 1 \Kkms{} may be described as
overly-bright in this way if \fH2\ = 0.5, hence the great majority of all
the statistically significant emission represented in Fig.  15 and in the
maps shown earlier for these sources.  For the brightest pixels
N(\HH)/\WCO\ $< 5\times10^{19} \HH\ \pcc $ (\Kkms)$^{-1}$.
 
The same \WCO-\EBV\ diagrams are shown for sources with higher \EBV\ in
Fig. 16.  Much of the gas around B2200+420 falls above the line for \fH2\ =
1, and attains such high brightness that its \HH/\WCO\ ratio is 3-4 times
below the standard conversion factor.  However, this case becomes
increasingly harder to make toward the other sources having higher \EBV\ as
in the bottom panels of Fig. 16.

\subsection{Sub-structure in reddening 
  would not eliminate large \WCO/\EBV\ ratios}

CO emission is heavily structured on arcminute scales, well below the
6\arcmin\ angular resolution of the reddening maps, and the high values and
large scatter in \WCO/\EBV\ in Fig. 15 cannot be accomodated with a fixed
ratio of \WCO/N(\HH) or \WCO/N(H) except by positing strong unresolved
variations, essentially clumping, in \EBV.  It is important to understand
the extent to which this might represent unresolved structure in the total
column density, for instance with regard to cleaning maps of the cosmic
microwave background \citep{bernard11}.  Given the extreme sensitivities of
the CO abundance and brightness to N(\HH) in diffuse clouds and the fact
that even \fH2\ may vary in diffuse material, it is entirely possible that
the large contrasts seen in \WCO\ do not have strong consequences for the
distribution of N(H), \EBV, or even N(\HH).

Shown in Fig. 17 are cumulative distribution functions of the integrated CO
emission \WCO\ in the fields around B0954+658 and \bll, using the native
ARO data and versions of the data smoothed to lower angular resolution
3\arcmin\ (similar to NANTEN) and 5\arcmin\ (similar to Planck).  The
brightness distribution of the CO around B0954+658 is compact in Fig. 3 but
still sufficiently extended that 4.5 \Kkms{} integrated intensities are
present at 5\arcmin\ resolution; this is well above the line for \fH2\ = 1
in Fig. 15.

The distribution of strongly emitting CO around \bll\ is sufficiently broad
in angle that 20-30\% of the pixels are occupied by CO with \WCO\ $\geq$ 5
K \kms\ whether the angular resolution is 1\arcmin\ or 5\arcmin; the very
strongest CO lines have \WCO\ $\ga$ 15\Kkms\ at 1-5\arcmin\ resolution in
the \bll\ field.  This is consistent with our recent observations of CO
emission in the field around \zoph\ \citep{LisPet+09} where the same peak
brightnesses were found in ARO and NANTEN data at 3\arcmin\ resolution.
 
\begin{figure}
  \includegraphics[height=8cm]{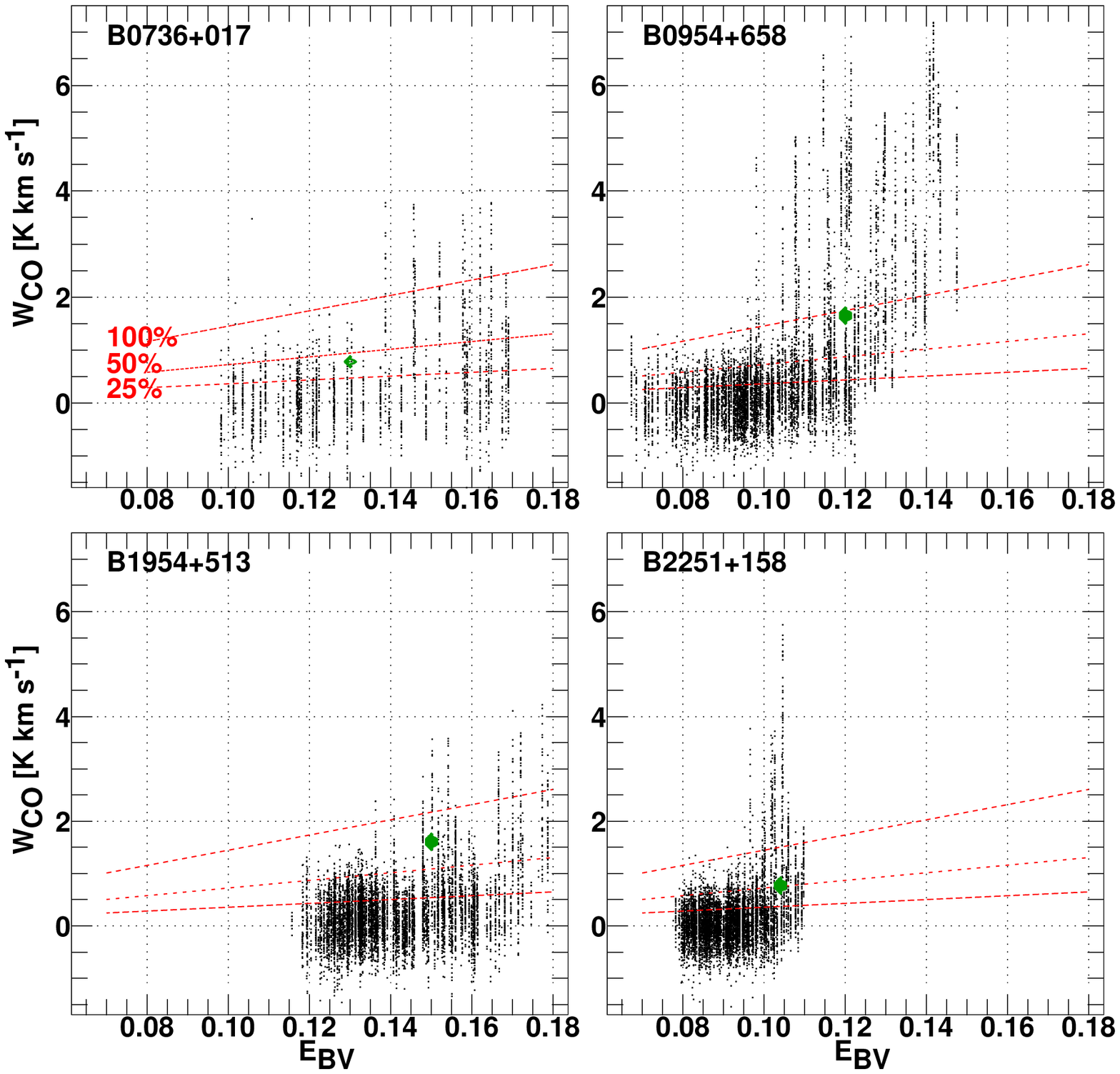}
\caption[]{ Distribution of \EBV\ and \WCO\ for four fields mapped here
  in CO.  Each 20\arcsec\ pixel in the CO maps is plotted as a separate
  point. The (red) dashed lines in each panel show the CO emission expected
  if 25\%, 50\% and 100\% of the gas is in molecular form with a typical
  value of the \WCO-N(\HH) conversion factor, N(\HH)/\WCO\ =
  $2\times10^{20}$\HH$\pcc$ (\Kkms)$^{-1}$.  In each panel a (green) filled
  diamond is shown at the value given in Table 1 toward the background
  source.}
\end{figure}

\begin{figure}
  \includegraphics[height=8cm]{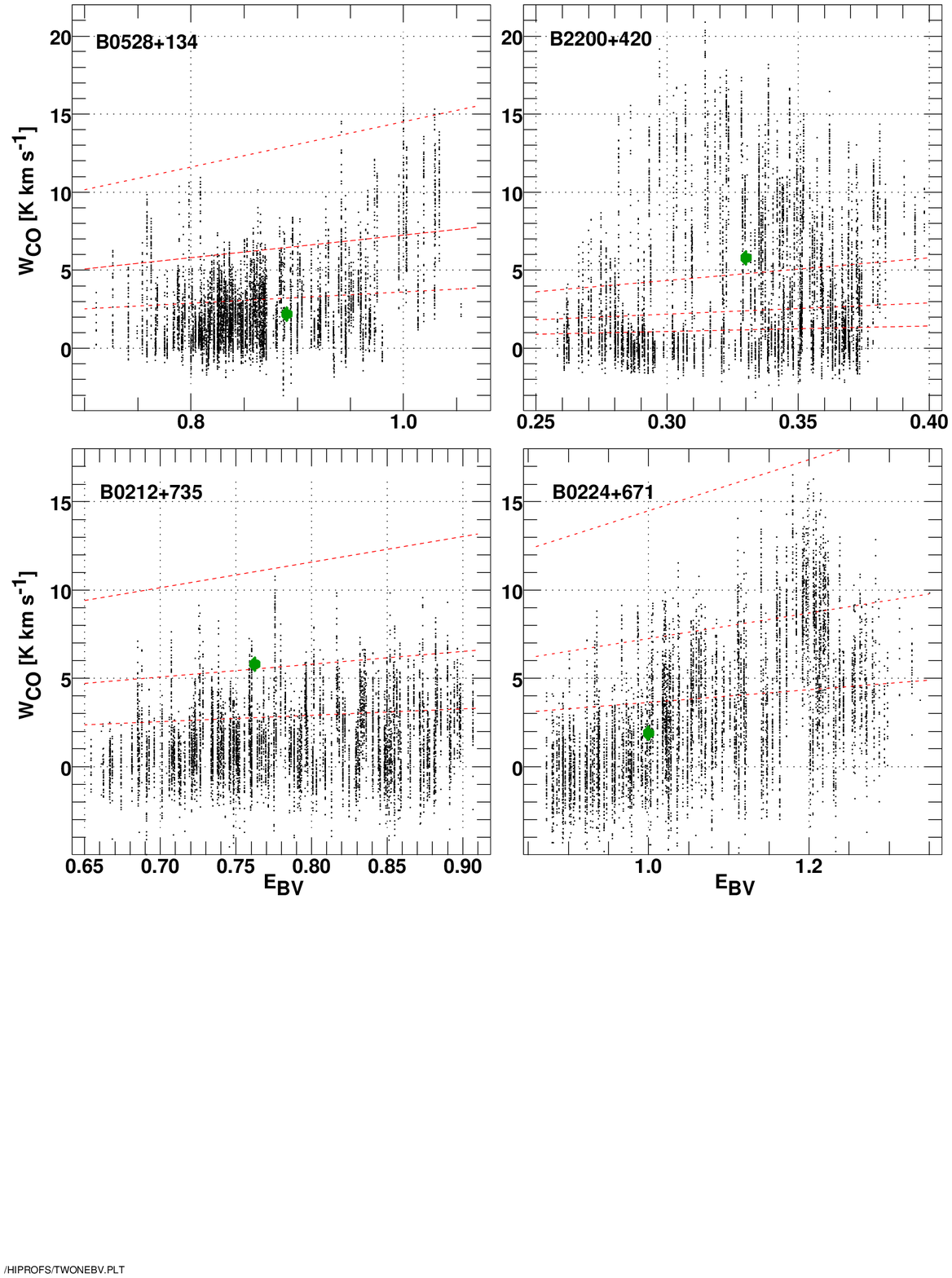}
\caption[]{As in Fig. 15 for four fields with larger reddening.}
\end{figure}

\begin{figure}
  \includegraphics[height=8cm]{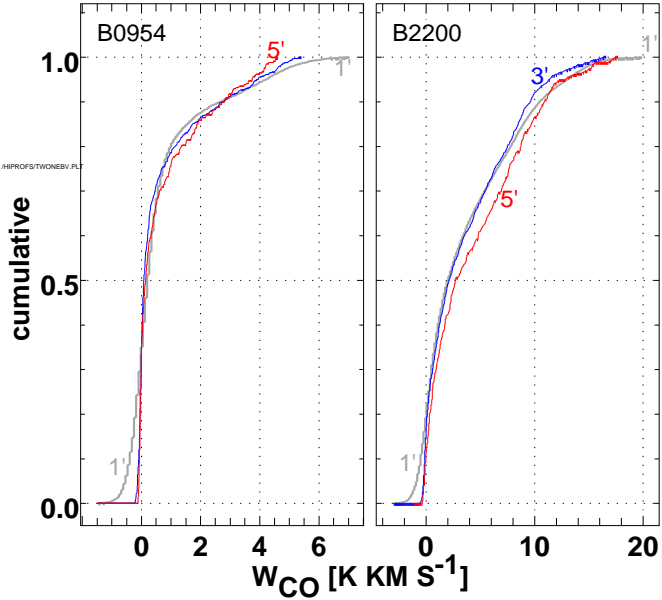}
\caption[]{Cumulative distribution functions for \WCO\ at spatial 
  resolutions of 1\arcmin\ (gray, thicker), 3\arcmin\ (blue) and 5\arcmin\ 
  (red) for B0954 (left) and B2200.}
\end{figure}

Because high CO brightness, and, therefore, impossibly high ratios
\WCO/\EBV\ (requiring \fH2\ $> 1$ for the mean \WCO/N(\HH) ratio) persist
to low angular resolution, \WCO\ cannot maintain a constant proportionality
to N(\HH) over the body of these clouds.  The observed variations in \WCO\ 
are too large to be accomodated by the total amount of material that is
present along the line of sight and unresolved structure in reddening
cannot account for the high values of \WCO/\EBV\ or the range of variation
in that ratio.

\subsection{Covering factors and bright and dark CO}

In Sect. 2.7 we noted the statistical certainty with which \hcop\ 
absorption is found in spectra taken within about 15\degr{} - 18\degr{} of
the galactic plane.  A corollary to this is that molecular gas is certain
to be omnipresent over the sky fields mapped in CO at such latitudes, no
matter how much of sky we actually found to be occupied by CO emission.

Table 2 shows pixel statistics for the CO emission maps made in the course
of this work; shown are profile channel-channel rms noise values and map
pixel-pixel rms noise in \WCO\ for each mappable kinematic component.  In
each case the covering factor was determined by forming a histogram of the
\WCO\ values and subtracting a gaussian fit to the component corresponding
to the noise, and is apparent in that it extends to unphysical negative
values of \WCO.  This is not a large correction; if the noise in \WCO\ is
random at a level $\sigma_{\rm map}$ the covering factor at/above any given
\WCO\ in the absence of signal is dA(\WCO)/A =
0.5(1-erf(\WCO/$\sqrt(2)\sigma_{\rm map}$)).  The expected covering factor
due to noise at even 2$\sigma$ significance is already below 3\%.

Table 2 shows that typical covering factors are 20\%-40\%, with a few null
values and only two sky fields (B0528 and B2200) where the majority of the
map area is covered.  Very approximately, the covering factors are small to
about the same extent that the \WCO/N(\HH) conversion factors of the
detected CO emission are higher than indicated by the standard CO-\HH\ 
conversion. In the end, some form of spatial averaging over brighter and
dimmer CO must be responsible for the global mean CO-\HH\ conversion factor
whether in fully molecular or diffuse molecular gas.  The idea that all gas
parcels would show identically the same \WCO/N(\HH) is preposterous.

\subsection{Pressure and density in CO-emitting gas}

Partial pressures of molecular hydrogen p(\HH)/k = n(\HH)\TK\ were derived
by \cite{LisLuc98} for all of the CO-bearing clouds discussed here, in the
directions of the background targets.  They generally fall in the range
$1-5 \times 10^{3} \pccc$ K, typical of thermal pressures in the diffuse
ISM sampled by neutral atomic carbon \citep{JenTri11} , which should share
the same volume. The CO derivation depends on recognizing that, when the
gas is warm and excitation is strongly sub-thermal, the excitation
temperature of the J=1-0 transition depends only on p(\HH) and the optical
depth of the transition $\tau(1-0)$.  In the limit of zero optical depth
the excitation temperature of the J=1-0 transition \Texc(1-0) is only
proportional to p(\HH), not to either the temperature or density
independently, making CO a useful barometer. This first became apparent in
the work of \cite{SmiSte+78} and is illustrated in Fig. 11 of
\cite{LisLuc98}: it has persisted over several generations of improved
collision cross sections.  Moreover, the excitation contribution from
atomic hydrogen should be small in CO-bearing gas where the molecular
fraction must be appreciable even if \HH\ does not dominate the overall
atomic-molecular balance \citep{Lis07CO}.

For $\tau(1-0) \la 3$ and p(\HH)/k $ \la 2\times 10^{4} \pccc$ K the
behaviour seen in Fig. 11 of \cite{LisLuc98} may be straightforwardly
parameterized to an accuracy of a few percent as

\begin{equation}
\Texc(1-0)-\Tcmb\ = 0.303 K \left[ \frac{p\left(\HH\right)}{10^3 \pccc~K} \right]^{1.02} 
  e^{\tau(1-0)^{0.6}/2.6}
\end{equation}

As examples of the application of this notion, we note:

\begin{itemize}
  
\item For a typical line with $\tau(1-0) = 1.5$ and a Rayleigh-Jeans
  brightness above the CMB of 1.5 K, \Texc(1-0) = 5.04 K and p(\HH)/k $=
  5\times 10^3 \pccc$ K or n(\HH) $= 200 \pccc$ at \TK\ = 25 K.
  
\item For the strongest absorption line component toward B0355 (-17.8
  \kms), $\tau(1-0) = 3.1$ and \Texc(1-0) $\approx 6$ K, so that p(\HH)/k
  $\approx\ 5\times 10^3 \pccc$ K once more.
  
\item The $\approx$ 4.5 K lines observed at the peak positions in several
  of the simple fields discussed in Section 3 require p(\HH)/k $\ga
  8.5\times 10^3 \pccc$ K or $\ga 15\times 10^3 \pccc$ K for $\tau(1-0)$ =
  3 or 1, respectively.  Such a heavy over-pressure must be transient.
  
\item The very bright 10-12 K lines seen near B0528+134 and B2200+420
  require excitation temperatures of 15 K or more and lie somewhat beyond
  the range where \WCO\ and N(CO) can be shown to be linearly proportional.
  They will be discussed separately in a forthcoming publication based on
  observations of \hcop, HCN, and CS.

\end{itemize}

\subsection{Failing to detect \HH\ when CO emission is weak}

There are cases where the brightness of the 1-0 line is well below 1 K even
when the CO optical depth is appreciable, as summarized in Table 3;
unfortunately we do not have a CO absorption profile toward B1928+738 in
whose field CO emission was not detected, see Sect. 3.4.  The regions of
very low p(\HH) toward B0212 and B0224 somehow manage to produce
appreciable amounts of CO without exciting it to detectable levels but
other lines represented in Table 3 do not arise in regions of especially
low pressure.

In Sect. 2.7 we showed that, on the whole, molecular gas is not
underrepresented by CO emission in the collection of sightlines comprising
this work and earlier we showed that the same is true of the larger sample
of absorption-cloud sightlines from which the current sample was drawn
\citep{LisPet+10}.  Moreover, CO emission from all of the components
represented in Table 3 is detected (usually quite strongly) elsewhere in
the mapped fields except around B1928.  However, the fraction of molecular
gas that is detectable in CO along individual sightlines varies
substantially.  To quantify this we equate the molecular column density
with the integrated optical depth measured in \hcop\ \citep{LucLis96}, see
the right-most column in Table 3.  In this case the fraction of molecular
gas that is missed by failing to detect CO emission from particular
individual components in four directions is 12\% toward B0212, 16\% toward
B0224, 8\% toward B0528+134 and 100\% toward B1928.  Overall the fraction
of molecular gas represented by the weakly-emitting CO summarized in Table
3 is 8\% toward B0528, 16\% toward B0224, 22\% toward B0212, 43\% toward
B0355 and 100\% toward B1730 and B1928.

\begin{table*}
\caption[]{Components with weak CO emission toward the continuum target$^a$}
{
\small
\begin{tabular}{lcccccccc}
\hline
Target & V   & $\tau(1-0)$ & \Tstar & dN(CO)/dV & \Texc(1-0) & p(\HH) & n(\HH)$^b$ & $\int \tau(\hcop)dv$/total \\ 
       & \kms  &        &   K  & $10^{15} \pcc$ (\kms)$^{-1}$  & K     & $10^3 \pccc$ K & $\pccc$ & \\
\hline
B0212& -10.3  & 0.49 & 0.40         & 0.65 &  3.4-3.6  & 1.9-2.5  & 75-100 &  0.122\\
     &  -0.05 & 0.95 & $<$ 0.14$^c$ & 1.11 & $<$3.1-3.2   & $<$0.9-1.1  & $<$35-45 & 0.102 \\
B0224 & -10.6 & 0.43 & $<$ 0.10     & 0.48 & $<$3.0-3.1 &  $<$0.8-1.0  & $<$30-40 & 0.161 \\
B0355&  -13.9 & 0.45 & 0.31         & 0.52 & 3.8-4.0 & 3.0-3.5  & 120-140 & 0.224\\
     &   -4.0 & 0.86    & 0.37         & 1.10  & 3.6-3.8 & 2.1-2.5  & 85-100 & 0.204\\
B0528&    2.8 & $<$ 0.11 & $<$ 0.16  & &  & &  & 0.080 \\
B1730&    5.1 & 1.15 & 0.24         & 1.42 & 3.5-3.7 & 1.8-2.2& 70-90 & 1.000\\
B1928&    $-3$ & $<$ 0.11 & $<$ 0.11  & &  & &  & 1.000 \\
\hline
\end{tabular}}
\\
$^a$ Using CO parameters originally derived by \cite{LisLuc98} and $\tau(\hcop)$ from \cite{LucLis96} \\
$^b$ At \TK\ = 25 K \\
$^c$ Upper limits in this column are $2\sigma$ \\

\end{table*}

\section{Discussion}

Even at \EBV\ = 0.1 - 0.3 mag, the CO emission traced in this work runs the
full gamut from undetectable to having brightness comparable to that seen
in fully-molecular dark clouds. CO emission may be undetectably weak ($<<$
1K) when molecular gas is present in absorption (including that of CO
itself) but in other directions it may be so bright that the N(\HH)/\WCO\ 
ratio is 4-5 times smaller than the typical CO-\HH\ conversion factor
$2\times10^{20}~\pcc$ (\Kkms)$^{-1}$.  Under the conditions encountered in
diffuse clouds, CO emission is foremost an indicator of the CO chemistry,
secondarily an indicator of the rotational excitation (which reflects the
partial thermal pressure of \HH) and only peripherally a measure of the
underlying hydrogen column density distribution as discussed in Sect. 8.
Indeed, the simulations of CO emission from the interstellar medium by
\citet{shetty11} agree with observations for the dense gas. However, a
detailed comparison with our results on the diffuse material shows that the
radiative transfer factor is correct but there are up to 4 orders of
magnitude difference in N(\HH)/N(CO). This is linked to the
poorly-understood polyatomic chemistry in the diffuse gas~\citep[see][their
section 4.3]{shetty11}.

The over-arching issues most relevant to diffuse cloud CO emission are
three-fold: 1) How it may be identified for what it is, originating in
relatively tenuous gas that is unassociated with star formation; 2) Whether
it makes a substantial contribution when CO emission is used as a surrogate
for N(\HH) in circumstances where emission contributions from diffuse and
dense heavily-shielded gas may be blended; 3) How it is related to the
so-called ``dark'' gas discovered by and FERMI \citep{grenier10} and PLANCK
\citep{bernard11} that is most prominent at moderate extinction where the
transition from atomic to molecular gas occurs and is claimed to host
50\%-120\% of the previously-known CO emitting gas in the solar
neighbourhood.

As for the identification of diffuse gas, the \WCO/\W13\ ratio is the most
acessible and direct probe.  When diffuse cloud CO is excited to detectable
levels it is generally in the regime where \WCO\ $\propto$ N(CO) and \W13\ 
$\propto$ N(\coth) so that the brightness ratio \WCO/\W13\ will be much
larger than the values 3-5 that are seen when emission arises from
optically thick lines from denser gas where the rotation ladder is close to
being thermalized.  Fractionation progressively lowers the
N(\cotw)/N(\coth) column density ratio in diffuse gas at larger N(CO)
\citep{LisLuc98,SheRog+07} but not below about 15.  Intensity ratios
\WCO/\W13\ of 8-10 or higher are a strong indicator that there is a major
contribution from diffuse material.

Regarding the contribution of diffuse gas we recently assessed it in the
case where an outside observer looked down on the Milky Way disk in the
vicinity of the Sun \citep{LisPet+10}.  We compared the mean emission for
the ensemble of lines of sight from which the current background targets
were drawn with the vertically-integrated emission expected for the
galactic disk component at the Solar Circle drawn from galactic plane
surveys.  The ensemble mean in our dataset, expressed as an equivalent to
looking vertically through the galactic layer, was 2$<$ \WCO\ sin($|$b$|$)
$>$ = 0.47 \Kkms{}.  The galactic disk contribution was inferred from
galactic plane surveys that find A(CO) = 5 \Kkms{} (kpc)$^{-1}$ and an
equivalent disk thickness of 150 pc, implying an integrated intensity
through the disk of 5 \Kkms{} (kpc)$^{-1} \times 0.15$ kpc = 0.75 \Kkms.

Even if viewed as entirely distinct (because it originates at galactic
latitudes well above those typically sampled in galactic plane surveys) the
diffuse gas contribution to the total seen looking down on the Milky Way
from outside would be 0.47/(0.47+0.75) = 38\%, a surprisingly high fraction
given the supposed absence of molecular gas and CO emission at higher
galactic latitudes.  The alternative, that the diffuse gas is already
incorporated in galactic plane surveys, makes the majority of the gas
(0.47/0.75) in the galactic plane diffuse.  This is an even more radical
proposition, but is consistent with finding that the preponderance of the
molecular gas seen toward the heavily-extincted line of sight toward
B0355+508 at $b=-1.6$\degr\ is actually diffuse.
 
In fact, the extent of the diffuse and/or high-latitude contribution to the
local CO emission remains to be determined by wide-field CO surveys whose
detection limit is substantially better than 1 \Kkms\ and perhaps no worse
than even 0.25 \Kkms.  Assessing the contribution of diffuse gas at lower
latitudes awaits a wider examination of the character of the gas seen in
the galactic disk, but the contribution of diffuse molecular gas in the
inner galactic disk is apparent in recent HERSCHEL/PRISMAS observations of
sub-mm absorption spectra toward star-forming regions
\citep{GerdeL+10,SonNeu+10}.

\section{Summary and conclusions}

We compared maps of CO emission with reddening maps on a typical field of
view of about $30'\times30'$ at an angular resolution of $1'$ toward 11
diffuse lines of sight for which we already had sub-arcsec molecular and/or
atomic absorption profiles. This allowed us to draw three kinds of
conclusions.

\subsection{ Conclusions about the position-position-velocity
  structure of the emission}
\begin{itemize}
  
\item Although most of the CO emission structure was amorphous or merely
  blob-like when mapped, the emission around B0528+138 was found to be
  highly regular and quasi-periodic while that around B2200+420 (aka \bll)
  was seen to be filamentary and tangled.
  
\item Toward B0355+508 and B0528+134, CO mapping suggests that pairs of
  absorption lines separated by 6-8 \kms\ are physically related, not
  merely accidental superpositions.
  
\item CO mapping shows that partition of an absorption profile into
  kinematic components, no matter how seemingly obvious, may actually be
  arbitrary and capricious: the decomposition may have no apparent validity
  in emission at positions only slightly removed from the continuum
  background.

\end{itemize}

\subsection{Conclusions linking the absorption to the
  emission kinematics:}
\begin{itemize}
  
\item The same clouds were seen in absorption and emission, and in atomic
  and molecular phases, although not necessarily in the same location.  We
  failed to find CO emission corresponding to just one out of 20 molecular
  absorption features, in one relatively small spatial field, i.e
  20\arcmin\ $\times$ 20\arcmin\ vs. 30\arcmin\ $\times$ 30\arcmin\ or
  more.  Conversely, while mapping away from the continuum background we
  saw only 2 CO emission features lacking molecular absorption
  counterparts.
  
\item CO emission was sometimes found in the field at velocities
  corresponding to features seen only in H I absorption toward the
  continuum.  We saw no molecular features outside the span of the H I
  absorption.
  
\end{itemize}

\subsection{Conclusions regarding the CO luminosity of diffuse gas.}
\begin{itemize}
  
\item We found relatively bright CO emission at modest reddening in the
  fields we mapped, with peak brightnesses of 4-5 K at \EBV\ $\la 0.15$ mag
  and up to 10-12 K at \EBV\ $\simeq$ 0.3 mag (i.e \AV $\simeq 1$ mag).
  This was true even for features that were seen only in absorption toward
  the continuum source in the field center.
  
\item The CO emission lines represent small column densities N(CO) $\leq
  10^{16} \pcc$, less than 10\% of the amount of free gas phase carbon
  expected along a line of sight with \EBV\ = 0.15 mag or \AV\ = 0.5 mag.
  The dominant form of gas phase carbon is still C\p.
  
\item When CO emission was detected at levels of 1.5 \Kkms{} and higher, it
  was generally over-luminous in the sense of having a small ratio
  N(\HH)/\WCO, i.e.  a value of the CO-\HH\ conversion factor below
  $2\times 10^{20}$\HH\ (\Kkms)$^{-1}$.  \WCO/N(\HH) ratios as small as
  N(\HH)/\WCO\ $\la 5\times 10^{19} \pcc$ \HH\ (\kms)$^{-1}$ are mandated
  by the observed reddening in cases where the line of sight was relatively
  free of extraneous material.
  
\item On the whole, the \WCO/N(\HH) ratio in diffuse gas is the same as in
  dense fully molecular clouds despite the presence of strong variations
  between individual diffuse gas parcels or sightlines.  The global
  \WCO/N(\HH) ratio in diffuse gas is the result of averaging over limited
  regions where CO emission is readily detectable and overly bright (in the
  sense of having \WCO/N(\HH) much higher than the mean), and with other
  regions having a significant molecular component (as seen in absorption)
  but where CO emission is comparatively weak or simply undetectable.
  
\item Small \WCO/N(\HH) ratios and sharp variations in the \WCO/\EBV\ ratio
  are not artifacts of the disparity in resolution between the 1\arcmin\ CO
  emission beam and the 6\arcmin\ resolution of the reddening maps, because
  high CO brightnesses and small \WCO/\EBV\ ratios persist when the
  resolution of the CO maps is degraded to that of the reddening maps.
  
\item Sharp variations in the CO emission brightness on arcminute scales do
  not necessarily represent unresolved structure in the reddening maps or
  in the column density of H or \HH.  Detectable CO emission generally
  arises in the regime where \WCO\ $\propto$ N(CO), and variations in the
  line brightness represent primarily the CO chemistry with its extreme
  sensitivity to \EBV\ and N(\HH). Secondarily the line brightness is
  influenced by CO rotational excitation since some features are not seen
  in emission toward continuum sources where there is CO absorption with
  appreciable optical depth.
  
\item Only peripherally does the CO brightness represent the underlying
  mass or \HH\ column density distribution of diffuse molecular gas.

\end{itemize}

\begin{acknowledgements}
  The National Radio Astronomy Observatory is operated by Associated
  Universites, Inc.  under a cooperative agreement with the US National
  Science Foundation.  The Kitt Peak 12-m millimetre wave telescope is
  operated by the Arizona Radio Observatory (ARO), Steward Observatory,
  University of Arizona.  IRAM is operated by CNRS (France), the MPG
  (Germany) and the IGN (Spain).  This work has been partially funded by
  the grant ANR-09-BLAN-0231-01 from the French {\it Agence Nationale de la
    Recherche} as part of the SCHISM project (http://schism.ens.fr/).  We
  thank Edith Falgarone for comments that inspired Sections 8.4 and 8.5 of
  this work.
\end{acknowledgements}

\Online{}

\begin{appendix}

\section{CO line kinematics around 3 additional objects}

Shown in Figs. A.1 - A.3 are position-velocity diagrams in right ascension
across the positions of B2251+ 158, b0212+735 and B0224+671.  As in Figs.
13 and 14 in the main text, the \hcop\ absorption spectrum toward the
continuum background target is superposed in the figures with its baseline
positioned where the diagram most nearly crosses the location of the
continuum.  These diagrams are intended to show how the features that occur
in the absorption line profiles are somewhat haphazard samples of the
larger scale gas distribution traced in CO emission but toward B0212+735
the diagram also indicates how the CO emission underrepresents the
molecular gas distribution.

\begin{figure}
  \includegraphics[height=6.7cm]{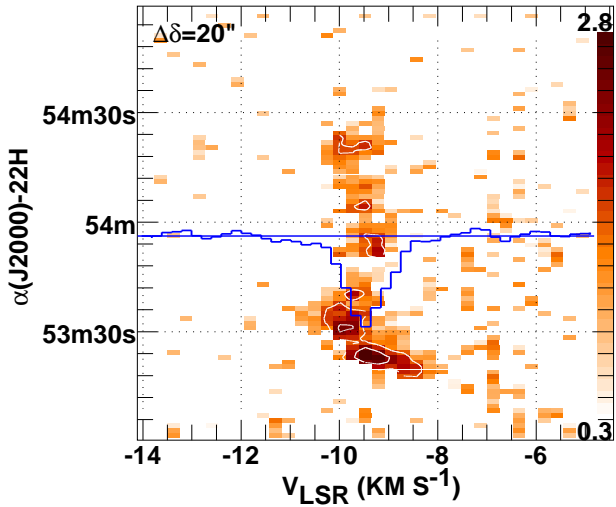}
\caption[]{A right ascension-velocity diagram of CO emission 20\arcsec\
  North of B2251+158.  The CO absorption spectrum toward B2251 is shown
  with its 0-level at the location of the continuum source; the peak
  absorption is 22\%, see Fig. 7.  Contours are shown at levels 1, 2, 3,
  ... K.}
\end{figure}

\begin{figure}
  \includegraphics[height=8.75cm]{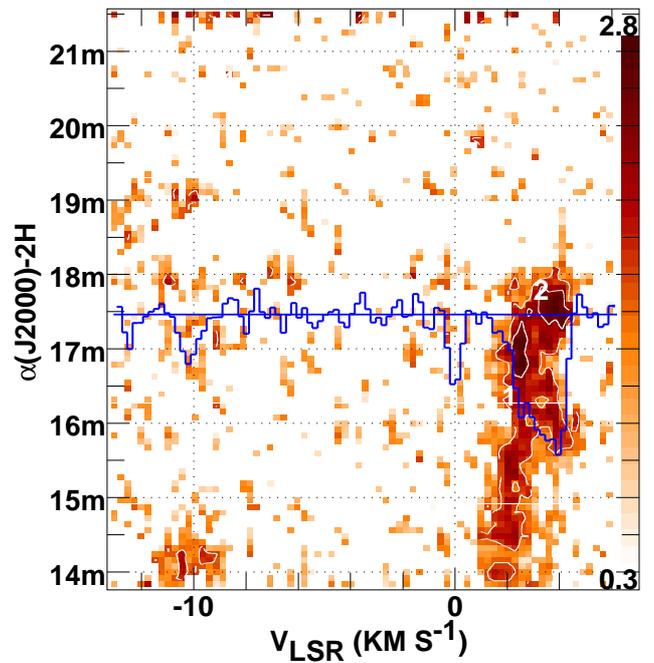}
\caption[]{A right ascension-velocity diagram  of CO emission across the field of 
  B0212+735.  The CO absorption spectrum toward B0212 is shown with its
  0-level at the location of the continuum source; the strongest absorption
  line is quite opaque, see Fig. 10.  Contours are shown at levels 1, 2, 3,
  ... K.}
\end{figure}

\begin{figure}
  \includegraphics[height=8.75cm]{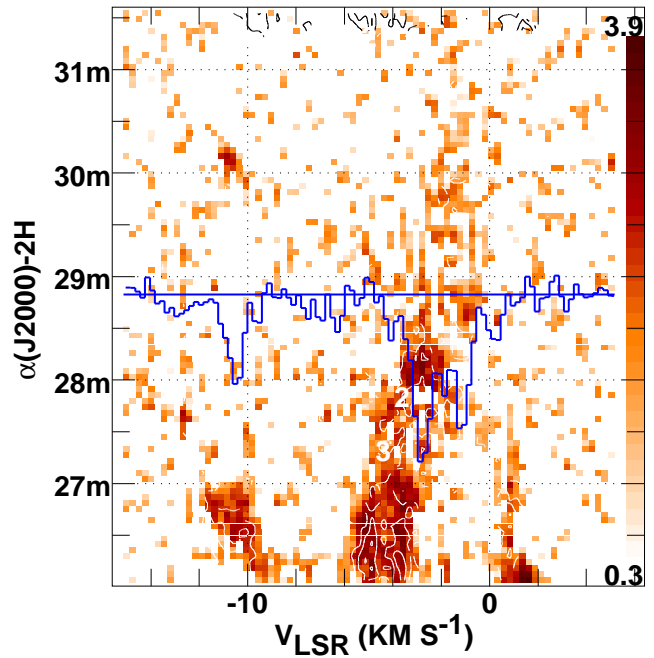}
\caption[]{A right ascension-velocity diagram of CO emission 
  across the field of B0224+671 at the declination of the continuum source.
  The CO absorption spectrum toward B0224+671 is shown with its 0-level at
  the location of the continuum source; the peak optical depth is 1.2, see
  Fig. 11.  Contours are shown at levels 1, 2, 3, ... K.}
\end{figure}

\section{The galaxy viewed in atomic gas around the background targets}

\begin{figure*}
  \includegraphics[height=22.5cm]{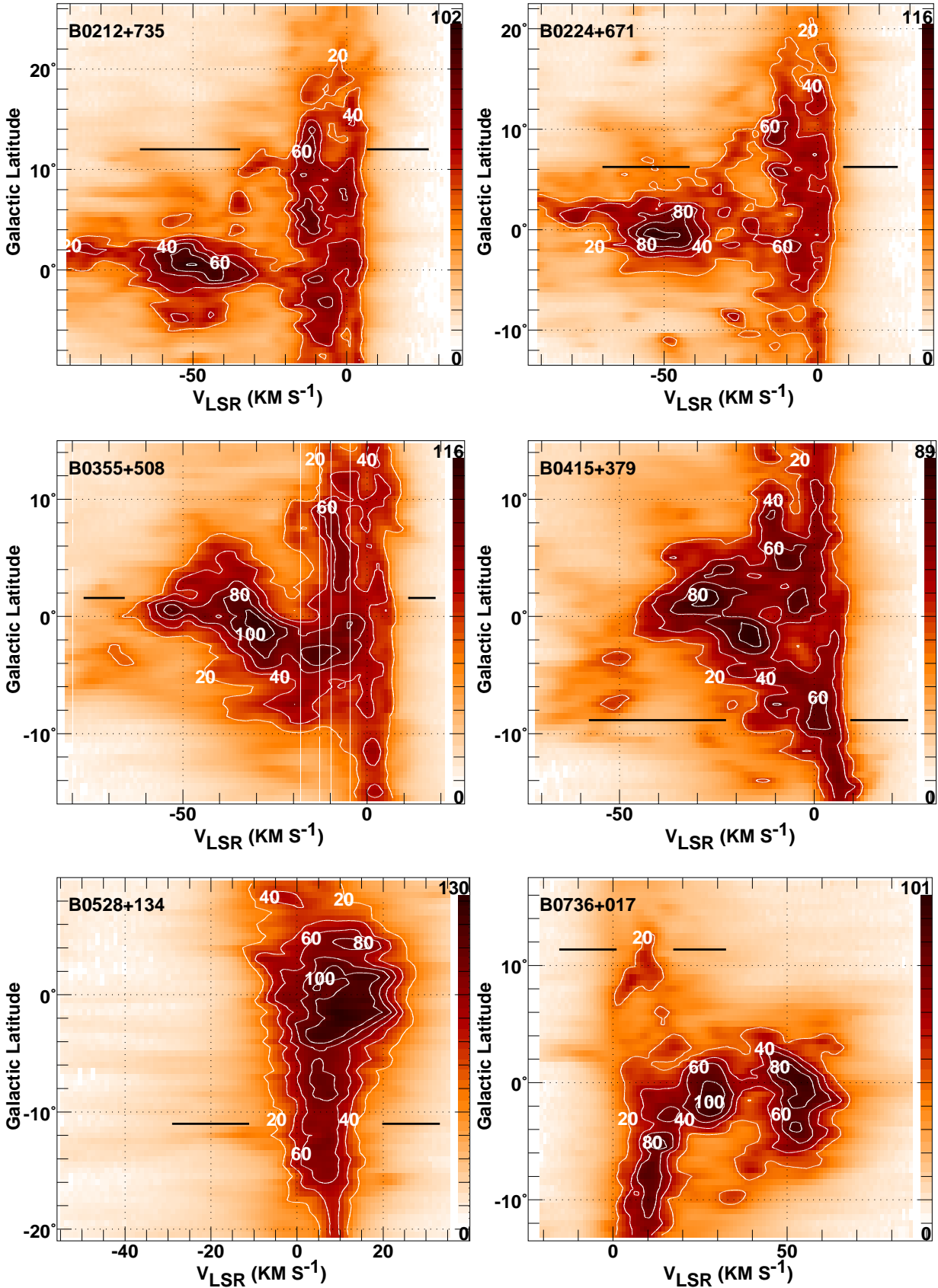}
\caption[]{Latitude-velocity diagrams of H I brightness around 
  six background targets used for mm-wave molecular absorption studies,
  using H I data from the LDSS survey \citep{HarBur97}.  The spatial
  resolution is 35\arcmin\ and the diagrams were constructed at the nearest
  longitude on the 0.5\degr\ grid of the survey datacube. The latitudes of
  the sources are marked.  The line of sight toward B0415+379 (3C111) is
  not discussed in this work.}
\end{figure*}

\begin{figure*}
  \includegraphics[height=22.5cm]{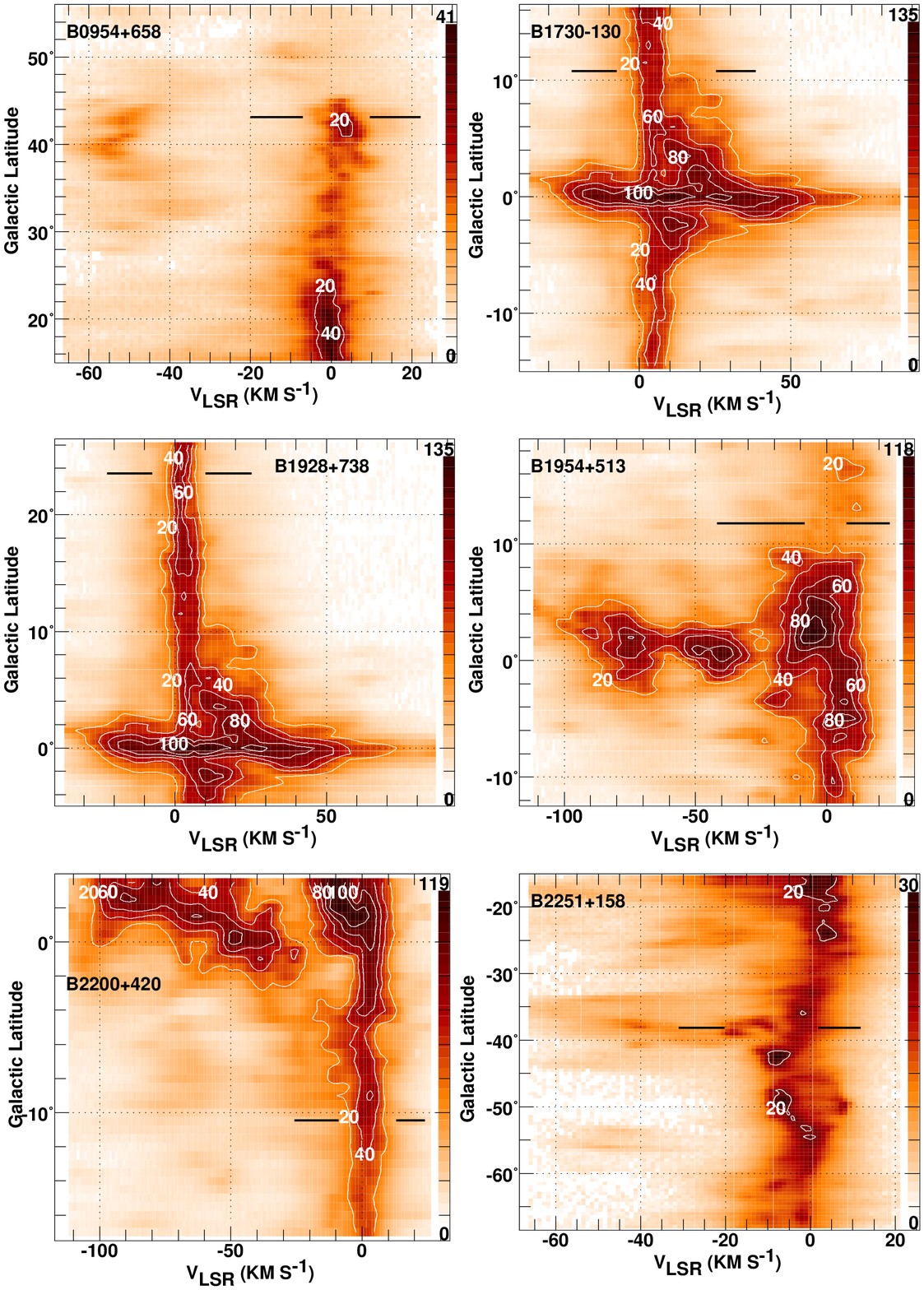}
\caption[]{Latitude-velocity diagrams of H I brightness around 
  six background targets used for mm-wave molecular absorption studies,
  using H I data from the LDSS survey \citep{HarBur97}, as in Fig. B.1 }
\end{figure*}

To provide context for the kinematics seen in the present work, Figs. B.1
and B.2 show large-galactic-scale H I latitude-velocity diagrams for each
of the sources studied here, using results of the LDSS survey of
\cite{HarBur97}.  The spatial resolution of these data is 35\arcmin\ and
the data are on a 30\arcmin\ grid.  The diagrams were constructed at the
galactic longitude nearest the background target (see Table 1).  The
latitudes of the sources are marked.  Even if it was not apparent from the
overlap of the H I and molecular absorption in the figures in the text,
these maps make it clear that the molecular gas studied here is part of
``normal'' galactic structure, mixed into the general ISM.

\end{appendix}


\begin{thebibliography}{47}
\expandafter\ifx\csname natexlab\endcsname\relax\def\natexlab#1{#1}\fi

\bibitem[{{Abdo} {et~al.}(2010){Abdo}, {Ackermann}, {Ajello}, {Baldini},
  {Ballet}, {Barbiellini}, {Bastieri}, {Baughman}, {Bechtol}, {Bellazzini},
  {Berenji}, {Bloom}, {Bonamente}, {Borgland}, {Bregeon}, {Brez}, {Brigida},
  {Bruel}, {Burnett}, {Buson}, {Caliandro}, {Cameron}, {Caraveo}, {Casandjian},
  {Cecchi}, {Celik}, {Chekhtman}, {Cheung}, {Chiang}, {Ciprini}, {Claus},
  {Cohen-Tanugi}, {Cominsky}, {Conrad}, {Dermer}, {de Palma}, {Digel}, {Silva},
  {Drell}, {Dubois}, {Dumora}, {Farnier}, {Favuzzi}, {Fegan}, {Focke},
  {Fortin}, {Frailis}, {Fukazawa}, {Funk}, {Fusco}, {Gargano}, {Gehrels},
  {Germani}, {Giavitto}, {Giebels}, {Giglietto}, {Giordano}, {Glanzman},
  {Godfrey}, {Grenier}, {Grondin}, {Grove}, {Guillemot}, {Guiriec}, {Harding},
  {Hayashida}, {Horan}, {Hughes}, {Jackson}, {J{\'o}hannesson}, {Johnson},
  {Johnson}, {Kamae}, {Katagiri}, {Kataoka}, {Kawai}, {Kerr}, {Kn{\"o}dlseder},
  {Kuss}, {Lande}, {Latronico}, {Lemoine-Goumard}, {Longo}, {Loparco}, {Lott},
  {Lovellette}, {Lubrano}, {Makeev}, {Mazziotta}, {McEnery}, {Meurer},
  {Michelson}, {Mitthumsiri}, {Mizuno}, {Monte}, {Monzani}, {Morselli},
  {Moskalenko}, {Murgia}, {Nolan}, {Norris}, {Nuss}, {Ohsugi}, {Okumura},
  {Omodei}, {Orlando}, {Ormes}, {Paneque}, {Pelassa}, {Pepe}, {Pesce-Rollins},
  {Piron}, {Porter}, {Rain{\`o}}, {Rando}, {Razzano}, {Reimer}, {Reimer},
  {Reposeur}, {Rodriguez}, {Ryde}, {Sadrozinski}, {Sanchez}, {Sander}, {Saz
  Parkinson}, {Sgr{\`o}}, {Siskind}, {Smith}, {Spandre}, {Spinelli}, {Starck},
  {Strickman}, {Strong}, {Suson}, {Takahashi}, {Tanaka}, {Thayer}, {Thayer},
  {Thompson}, {Tibaldo}, {Torres}, {Tosti}, {Tramacere}, {Uchiyama}, {Usher},
  {Vasileiou}, {Vilchez}, {Vitale}, {Waite}, {Wang}, {Winer}, {Wood}, {Ylinen},
  {Ziegler}, \& {Fermi/LAT Collaboration}}]{grenier10}
{Abdo}, A.~A., {Ackermann}, M., {Ajello}, M., {Baldini}, L., {Ballet}, J.,
  {Barbiellini}, G., {Bastieri}, D., {Baughman}, B.~M., {Bechtol}, K.,
  {Bellazzini}, R., {Berenji}, B., {Bloom}, E.~D., {Bonamente}, E., {Borgland},
  A.~W., {Bregeon}, J., {Brez}, A., {Brigida}, M., {Bruel}, P., {Burnett},
  T.~H., {Buson}, S., {Caliandro}, G.~A., {Cameron}, R.~A., {Caraveo}, P.~A.,
  {Casandjian}, J.~M., {Cecchi}, C., {Celik}, {\"O}., {Chekhtman}, A.,
  {Cheung}, C.~C., {Chiang}, J., {Ciprini}, S., {Claus}, R., {Cohen-Tanugi},
  J., {Cominsky}, L.~R., {Conrad}, J., {Dermer}, C.~D., {de Palma}, F.,
  {Digel}, S.~W., {Silva}, E.~d.~C.~e., {Drell}, P.~S., {Dubois}, R., {Dumora},
  D., {Farnier}, C., {Favuzzi}, C., {Fegan}, S.~J., {Focke}, W.~B., {Fortin},
  P., {Frailis}, M., {Fukazawa}, Y., {Funk}, S., {Fusco}, P., {Gargano}, F.,
  {Gehrels}, N., {Germani}, S., {Giavitto}, G., {Giebels}, B., {Giglietto}, N.,
  {Giordano}, F., {Glanzman}, T., {Godfrey}, G., {Grenier}, I.~A., {Grondin},
  M.-H., {Grove}, J.~E., {Guillemot}, L., {Guiriec}, S., {Harding}, A.~K.,
  {Hayashida}, M., {Horan}, D., {Hughes}, R.~E., {Jackson}, M.~S.,
  {J{\'o}hannesson}, G., {Johnson}, A.~S., {Johnson}, W.~N., {Kamae}, T.,
  {Katagiri}, H., {Kataoka}, J., {Kawai}, N., {Kerr}, M., {Kn{\"o}dlseder}, J.,
  {Kuss}, M., {Lande}, J., {Latronico}, L., {Lemoine-Goumard}, M., {Longo}, F.,
  {Loparco}, F., {Lott}, B., {Lovellette}, M.~N., {Lubrano}, P., {Makeev}, A.,
  {Mazziotta}, M.~N., {McEnery}, J.~E., {Meurer}, C., {Michelson}, P.~F.,
  {Mitthumsiri}, W., {Mizuno}, T., {Monte}, C., {Monzani}, M.~E., {Morselli},
  A., {Moskalenko}, I.~V., {Murgia}, S., {Nolan}, P.~L., {Norris}, J.~P.,
  {Nuss}, E., {Ohsugi}, T., {Okumura}, A., {Omodei}, N., {Orlando}, E.,
  {Ormes}, J.~F., {Paneque}, D., {Pelassa}, V., {Pepe}, M., {Pesce-Rollins},
  M., {Piron}, F., {Porter}, T.~A., {Rain{\`o}}, S., {Rando}, R., {Razzano},
  M., {Reimer}, A., {Reimer}, O., {Reposeur}, T., {Rodriguez}, A.~Y., {Ryde},
  F., {Sadrozinski}, H.~F.-W., {Sanchez}, D., {Sander}, A., {Saz Parkinson},
  P.~M., {Sgr{\`o}}, C., {Siskind}, E.~J., {Smith}, P.~D., {Spandre}, G.,
  {Spinelli}, P., {Starck}, J.-L., {Strickman}, M.~S., {Strong}, A.~W.,
  {Suson}, D.~J., {Takahashi}, H., {Tanaka}, T., {Thayer}, J.~B., {Thayer},
  J.~G., {Thompson}, D.~J., {Tibaldo}, L., {Torres}, D.~F., {Tosti}, G.,
  {Tramacere}, A., {Uchiyama}, Y., {Usher}, T.~L., {Vasileiou}, V., {Vilchez},
  N., {Vitale}, V., {Waite}, A.~P., {Wang}, P., {Winer}, B.~L., {Wood}, K.~S.,
  {Ylinen}, T., {Ziegler}, M., \& {Fermi/LAT Collaboration}. 2010, \apj, 710,
  133

\bibitem[{{Bania} {et~al.}(1991){Bania}, {Marscher}, \&
  {Barvainis}}]{BanMar+91}
{Bania}, T.~M., {Marscher}, A.~P., \& {Barvainis}, R. 1991, Astron. J., 101,
  2147

\bibitem[{{Bern{\'e}} {et~al.}(2010){Bern{\'e}}, {Marcelino}, \&
  {Cernicharo}}]{BerMar+10}
{Bern{\'e}}, O., {Marcelino}, N., \& {Cernicharo}, J. 2010, Nature, 466, 947

\bibitem[{{Burgh} {et~al.}(2007){Burgh}, {France}, \& {McCandliss}}]{BurFra+07}
{Burgh}, E.~B., {France}, K., \& {McCandliss}, S.~R. 2007, ApJ, 658, 446

\bibitem[{{Dame} {et~al.}(2001){Dame}, {Hartmann}, \& {Thaddeus}}]{DamHar+01}
{Dame}, T.~M., {Hartmann}, D., \& {Thaddeus}, P. 2001, ApJ, 547, 792

\bibitem[{{Dame} \& {Thaddeus}(1994)}]{DamTha94}
{Dame}, T.~M. \& {Thaddeus}, P. 1994, ApJ, 436, L173

\bibitem[{{Dickey} {et~al.}(1983){Dickey}, {Kulkarni}, {Heiles}, \& {Van
  Gorkom}}]{DicKul+83}
{Dickey}, J.~M., {Kulkarni}, S.~R., {Heiles}, C.~E., \& {Van Gorkom}, J.~H.
  1983, Astrophys. J., Suppl. Ser., 53, 591

\bibitem[{{Gerin} {et~al.}(2010){Gerin}, {de Luca}, {Goicoechea}, {Herbst},
  {Falgarone}, {Godard}, {Bell}, {Coutens}, {Ka{\'z}mierczak}, {Sonnentrucker},
  {Black}, {Neufeld}, {Phillips}, {Pearson}, {Rimmer}, {Hassel}, {Lis},
  {Vastel}, {Boulanger}, {Cernicharo}, {Dartois}, {Encrenaz}, {Giesen},
  {Goldsmith}, {Gupta}, {Gry}, {Hennebelle}, {Hily-Blant}, {Joblin},
  {Ko{\l}os}, {Kre{\l}owski}, {Mart{\'{\i}}n-Pintado}, {Monje}, {Mookerjea},
  {Perault}, {Persson}, {Plume}, {Salez}, {Schmidt}, {Stutzki}, {Teyssier},
  {Yu}, {Contursi}, {Menten}, {Geballe}, {Schlemmer}, {Morris}, {Hatch},
  {Imram}, {Ward}, {Caux}, {G{\"u}sten}, {Klein}, {Roelfsema}, {Dieleman},
  {Schieder}, {Honingh}, \& {Zmuidzinas}}]{GerdeL+10}
{Gerin}, M., {de Luca}, M., {Goicoechea}, J.~R., {Herbst}, E., {Falgarone}, E.,
  {Godard}, B., {Bell}, T.~A., {Coutens}, A., {Ka{\'z}mierczak}, M.,
  {Sonnentrucker}, P., {Black}, J.~H., {Neufeld}, D.~A., {Phillips}, T.~G.,
  {Pearson}, J., {Rimmer}, P.~B., {Hassel}, G., {Lis}, D.~C., {Vastel}, C.,
  {Boulanger}, F., {Cernicharo}, J., {Dartois}, E., {Encrenaz}, P., {Giesen},
  T., {Goldsmith}, P.~F., {Gupta}, H., {Gry}, C., {Hennebelle}, P.,
  {Hily-Blant}, P., {Joblin}, C., {Ko{\l}os}, R., {Kre{\l}owski}, J.,
  {Mart{\'{\i}}n-Pintado}, J., {Monje}, R., {Mookerjea}, B., {Perault}, M.,
  {Persson}, C., {Plume}, R., {Salez}, M., {Schmidt}, M., {Stutzki}, J.,
  {Teyssier}, D., {Yu}, S., {Contursi}, A., {Menten}, K., {Geballe}, T.~R.,
  {Schlemmer}, S., {Morris}, P., {Hatch}, W.~A., {Imram}, M., {Ward}, J.~S.,
  {Caux}, E., {G{\"u}sten}, R., {Klein}, T., {Roelfsema}, P., {Dieleman}, P.,
  {Schieder}, R., {Honingh}, N., \& {Zmuidzinas}, J. 2010, A\&A, 521, L16

\bibitem[{{Gir} {et~al.}(1994){Gir}, {Blitz}, \& {Magnani}}]{GirBli+94}
{Gir}, B.-Y., {Blitz}, L., \& {Magnani}, L. 1994, ApJ, 434, 162

\bibitem[{{Goldreich} \& {Kwan}(1974)}]{GolKwa74}
{Goldreich}, P. \& {Kwan}, J. 1974, ApJ, 189, 441

\bibitem[{{Hartmann} \& {Burton}(1997)}]{HarBur97}
{Hartmann}, D. \& {Burton}, W.~B. 1997, Atlas of galactic neutral hydrogen
  (Cambridge; New York: Cambridge University Press)

\bibitem[{{Heithausen}(2004)}]{Hei04}
{Heithausen}, A. 2004, ApJ, 606, L13

\bibitem[{{Hily-Blant} \& {Falgarone}(2009)}]{HilFal09}
{Hily-Blant}, P. \& {Falgarone}, E. 2009, A\&A, 500, L29

\bibitem[{{Hogerheijde} {et~al.}(1995){Hogerheijde}, {De Geus}, \&
  {Spaans}}]{HogDeG+95}
{Hogerheijde}, M.~R., {De Geus}, E.~J., \& {Spaans}, F. 1995, ApJ, 441, L93

\bibitem[{{Jenkins} \& {Tripp}(2011)}]{JenTri11}
{Jenkins}, E.~B. \& {Tripp}, T.~M. 2011, ApJ, 734, 65

\bibitem[{{Lequeux} {et~al.}(1993){Lequeux}, {Allen}, \&
  {Guilloteau}}]{LeqAll+93}
{Lequeux}, J., {Allen}, R.~J., \& {Guilloteau}, S. 1993, A\&A, 280, L23

\bibitem[{{Liszt} \& {Lucas}(2001)}]{LisLuc01}
{Liszt}, H. \& {Lucas}, R. 2001, A\&A, 370, 576

\bibitem[{{Liszt} {et~al.}(2006){Liszt}, {Lucas}, \& {Pety}}]{LisLuc+06}
{Liszt}, H., {Lucas}, R., \& {Pety}, J. 2006, A\&A, 448, 253

\bibitem[{{Liszt}(1994)}]{Lis94}
{Liszt}, H.~S. 1994, ApJ, 429, 638

\bibitem[{{Liszt}(2007)}]{Lis07CO}
---. 2007, A\&A, 476, 291

\bibitem[{{Liszt} \& {Lucas}(1996)}]{LisLuc96}
{Liszt}, H.~S. \& {Lucas}, R. 1996, A\&A, 314, 917

\bibitem[{{Liszt} \& {Lucas}(1998)}]{LisLuc98}
---. 1998, A\&A, 339, 561

\bibitem[{{Liszt} \& {Lucas}(2000)}]{LisLuc00}
---. 2000, A\&A, 355, 333

\bibitem[{{Liszt} {et~al.}(2010){Liszt}, {Pety}, \& {Lucas}}]{LisPet+10}
{Liszt}, H.~S., {Pety}, J., \& {Lucas}, R. 2010, A\&A, 518, A45+

\bibitem[{{Liszt} {et~al.}(2009){Liszt}, {Pety}, \& {Tachihara}}]{LisPet+09}
{Liszt}, H.~S., {Pety}, J., \& {Tachihara}, K. 2009, A\&A, 499, 503

\bibitem[{{Liszt} \& {Wilson}(1993)}]{LisWil93}
{Liszt}, H.~S. \& {Wilson}, R.~W. 1993, ApJ, 403, 663

\bibitem[{{Lucas} \& {Liszt}(1993)}]{LucLis93}
{Lucas}, R. \& {Liszt}, H.~S. 1993, A\&A, 276, L33

\bibitem[{{Lucas} \& {Liszt}(1996)}]{LucLis96}
---. 1996, A\&A, 307, 237

\bibitem[{{Lucas} \& {Liszt}(2000)}]{LucLis00C2H}
---. 2000, A\&A, 358, 1069

\bibitem[{{Maddalena} \& {Morris}(1987)}]{MadMor87}
{Maddalena}, R.~J. \& {Morris}, M. 1987, ApJ, 323, 179

\bibitem[{{Magnani} {et~al.}(1985){Magnani}, {Blitz}, \& {Mundy}}]{MagBli+85}
{Magnani}, L., {Blitz}, L., \& {Mundy}, L. 1985, ApJ, 295, 402

\bibitem[{{Marscher} {et~al.}(1991){Marscher}, {Bania}, \& {Wang}}]{MarBan+91}
{Marscher}, A.~P., {Bania}, T.~M., \& {Wang}, Z. 1991, ApJ, 371, L77

\bibitem[{{Pety} \& {Falgarone}(2003)}]{PetFal03}
{Pety}, J. \& {Falgarone}, E. 2003, A\&A, 412, 417

\bibitem[{{Pety} {et~al.}(2008){Pety}, {Lucas}, \& {Liszt}}]{PetLuc+08}
{Pety}, J., {Lucas}, R., \& {Liszt}, H.~S. 2008, A\&A, 489, 217

\bibitem[{{Planck Collaboration} {et~al.}(2011){Planck Collaboration}, {Ade},
  {Aghanim}, {Arnaud}, {Ashdown}, {Aumont}, {Baccigalupi}, {Balbi}, {Banday},
  {Barreiro}, \& et~al.}]{bernard11}
{Planck Collaboration}, {Ade}, P.~A.~R., {Aghanim}, N., {Arnaud}, M.,
  {Ashdown}, M., {Aumont}, J., {Baccigalupi}, C., {Balbi}, A., {Banday}, A.~J.,
  {Barreiro}, R.~B., \& et~al. 2011, \aap, 536, A19

\bibitem[{{Rachford} {et~al.}(2009){Rachford}, {Snow}, {Destree}, {Ross},
  {Ferlet}, {Friedman}, {Gry}, {Jenkins}, {Morton}, {Savage}, {Shull},
  {Sonnentrucker}, {Tumlinson}, {Vidal-Madjar}, {Welty}, \& {York}}]{RacSno+09}
{Rachford}, B.~L., {Snow}, T.~P., {Destree}, J.~D., {Ross}, T.~L., {Ferlet},
  R., {Friedman}, S.~D., {Gry}, C., {Jenkins}, E.~B., {Morton}, D.~C.,
  {Savage}, B.~D., {Shull}, J.~M., {Sonnentrucker}, P., {Tumlinson}, J.,
  {Vidal-Madjar}, A., {Welty}, D.~E., \& {York}, D.~G. 2009, Astrophys. J.,
  Suppl. Ser., 180, 125

\bibitem[{{Sakamoto} \& {Sunada}(2003)}]{SakSun03}
{Sakamoto}, S. \& {Sunada}, K. 2003, ApJ, 594, 340

\bibitem[{{Savage} {et~al.}(1977){Savage}, {Drake}, {Budich}, \&
  {Bohlin}}]{SavDra+77}
{Savage}, B.~D., {Drake}, J.~F., {Budich}, W., \& {Bohlin}, R.~C. 1977, ApJ,
  216, 291

\bibitem[{{Schlegel} {et~al.}(1998){Schlegel}, {Finkbeiner}, \&
  {Davis}}]{SchFin+98}
{Schlegel}, D.~J., {Finkbeiner}, D.~P., \& {Davis}, M. 1998, ApJ, 500, 525

\bibitem[{{Sheffer} {et~al.}(2008){Sheffer}, {Rogers}, {Federman}, {Abel},
  {Gredel}, {Lambert}, \& {Shaw}}]{SheRog+08}
{Sheffer}, Y., {Rogers}, M., {Federman}, S.~R., {Abel}, N.~P., {Gredel}, R.,
  {Lambert}, D.~L., \& {Shaw}, G. 2008, ApJ, 687, 1075

\bibitem[{{Sheffer} {et~al.}(2007){Sheffer}, {Rogers}, {Federman}, {Lambert},
  \& {Gredel}}]{SheRog+07}
{Sheffer}, Y., {Rogers}, M., {Federman}, S.~R., {Lambert}, D.~L., \& {Gredel},
  R. 2007, ApJ, 667, 1002

\bibitem[{{Shetty} {et~al.}(2011){Shetty}, {Glover}, {Dullemond}, \&
  {Klessen}}]{shetty11}
{Shetty}, R., {Glover}, S.~C., {Dullemond}, C.~P., \& {Klessen}, R.~S. 2011,
  \mnras, 412, 1686

\bibitem[{{Smith} {et~al.}(1978){Smith}, {Stecher}, \& {Krishna
  Swamy}}]{SmiSte+78}
{Smith}, A.~M., {Stecher}, T.~P., \& {Krishna Swamy}, K.~S. 1978, ApJ, 220, 138

\bibitem[{{Snow} \& {McCall}(2006)}]{SnoMcC06}
{Snow}, T.~P. \& {McCall}, B.~J. 2006, Ann. Rev. Astrophys. Astron., 44, 367

\bibitem[{{Sonnentrucker} {et~al.}(2010){Sonnentrucker}, {Neufeld}, {Phillips},
  {Gerin}, {Lis}, {de Luca}, {Goicoechea}, {Black}, {Bell}, {Boulanger},
  {Cernicharo}, {Coutens}, {Dartois}, {Ka{\'z}mierczak}, {Encrenaz},
  {Falgarone}, {Geballe}, {Giesen}, {Godard}, {Goldsmith}, {Gry}, {Gupta},
  {Hennebelle}, {Herbst}, {Hily-Blant}, {Joblin}, {Ko{\l}os}, {Kre{\l}owski},
  {Mart{\'{\i}}n-Pintado}, {Menten}, {Monje}, {Mookerjea}, {Pearson},
  {Perault}, {Persson}, {Plume}, {Salez}, {Schlemmer}, {Schmidt}, {Stutzki},
  {Teyssier}, {Vastel}, {Yu}, {Caux}, {G{\"u}sten}, {Hatch}, {Klein}, {Mehdi},
  {Morris}, \& {Ward}}]{SonNeu+10}
{Sonnentrucker}, P., {Neufeld}, D.~A., {Phillips}, T.~G., {Gerin}, M., {Lis},
  D.~C., {de Luca}, M., {Goicoechea}, J.~R., {Black}, J.~H., {Bell}, T.~A.,
  {Boulanger}, F., {Cernicharo}, J., {Coutens}, A., {Dartois}, E.,
  {Ka{\'z}mierczak}, M., {Encrenaz}, P., {Falgarone}, E., {Geballe}, T.~R.,
  {Giesen}, T., {Godard}, B., {Goldsmith}, P.~F., {Gry}, C., {Gupta}, H.,
  {Hennebelle}, P., {Herbst}, E., {Hily-Blant}, P., {Joblin}, C., {Ko{\l}os},
  R., {Kre{\l}owski}, J., {Mart{\'{\i}}n-Pintado}, J., {Menten}, K.~M.,
  {Monje}, R., {Mookerjea}, B., {Pearson}, J., {Perault}, M., {Persson}, C.~M.,
  {Plume}, R., {Salez}, M., {Schlemmer}, S., {Schmidt}, M., {Stutzki}, J.,
  {Teyssier}, D., {Vastel}, C., {Yu}, S., {Caux}, E., {G{\"u}sten}, R.,
  {Hatch}, W.~A., {Klein}, T., {Mehdi}, I., {Morris}, P., \& {Ward}, J.~S.
  2010, A\&A, 521, L12

\bibitem[{{Sonnentrucker} {et~al.}(2007){Sonnentrucker}, {Welty}, {Thorburn},
  \& {York}}]{SonWel+07}
{Sonnentrucker}, P., {Welty}, D.~E., {Thorburn}, J.~A., \& {York}, D.~G. 2007,
  Astrophys. J., Suppl. Ser., 168, 58

\bibitem[{{Yamamoto} {et~al.}(2003){Yamamoto}, {Onishi}, {Mizuno}, \&
  {Fukui}}]{YamOni+03}
{Yamamoto}, H., {Onishi}, T., {Mizuno}, A., \& {Fukui}, Y. 2003, ApJ, 592, 217

\end{thebibliography}
\end{document}